\documentclass[preprintnumbers,prd,superscriptaddress,preprint]{revtex4-1}

\usepackage{CJK}
\usepackage{amsmath}
\usepackage{subfigure}
\usepackage{enumerate}
\usepackage{amssymb}
\usepackage{amsfonts}
\usepackage{mathrsfs}
\usepackage{latexsym}
\usepackage{bm}
\usepackage{amscd}
\usepackage{physics} 
\usepackage{graphicx}
\usepackage{epsfig}
\usepackage{hhline,multirow}
\usepackage{dcolumn}
\usepackage{url}
\usepackage[colorlinks=true,linkcolor=blue,citecolor=blue,urlcolor=blue]{hyperref}
\usepackage{color}
\usepackage{soul}
\usepackage{xcolor}
\usepackage{indentfirst}
\usepackage{subfigure}
\usepackage{psfrag}
\usepackage{slashed}
\usepackage{float}
\usepackage{setspace}
\usepackage{empheq}
\usepackage{mathtools}
\allowdisplaybreaks[2]

\newcommand{\MBbar}{\overline{M}_{\!B}}
\newcommand{\MTbar}{\overline{M}_T}

\begin{document}

\preprint{JLAB-THY-24-4069}

\title{Nonlocal chiral contributions to generalized parton distributions of the proton at nonzero skewness}

\author{Zhengyang Gao}
\affiliation{Institute of High Energy Physics, Chinese Academy of Sciences,
	Beijing 100049, China}
\affiliation{\mbox{College of Physics Sciences, University of Chinese Academy of Sciences, Beijing 100049, China}}
\author{Fangcheng He}
\affiliation{Department of Physics and Astronomy, Stony Brook University, Stony Brook, New York 11790, USA}
\author{Chueng-Ryong~Ji}
\affiliation{{Department of Physics, North Carolina State University,
	Raleigh, North Carolina 27695, USA}}
\author{W. Melnitchouk}
\affiliation{Jefferson Lab, Newport News, Virginia 23606, USA}
\author{Y.~Salamu}
\affiliation{{School of Physics and Electrical Engineering, Kashi University, Kashi, 844000, Xinjiang, China}}
\author{P. Wang}
\affiliation{Institute of High Energy Physics, Chinese Academy of Sciences,
	Beijing 100049, China}
\affiliation{\mbox{College of Physics Sciences, University of Chinese Academy of Sciences, Beijing 100049, China}}

\begin{abstract}
We compute the one-loop contributions to spin-averaged generalized parton distributions (GPDs) in the proton from pseudoscalar mesons with intermediate octet and decuplet baryon states at nonzero skewness.
Our framework is based on nonlocal covariant chiral effective theory, with ultraviolet divergences regularized by introducing a relativistic regulator derived consistently from the nonlocal Lagrangian.
Using the splitting functions calculated from the nonlocal Lagrangian, we find the nonzero skewness GPDs from meson loops by convoluting with the phenomenological pion GPD and the generalized distribution amplitude, and verify that these satisfy the correct polynomiality properties.
We also compute the lowest two moments of GPDs to quantify the meson loop effects on the Dirac, Pauli and gravitational form factors of the proton.
\end{abstract}

\date{\today}
\maketitle

\section{Introduction}

Generalized parton distributions (GPDs) contain unique information about the 3-dimensional structure of hadrons, in terms of their fundamental quark and gluon (or parton) constituents.
They describe the distributions in momentum and position space of partons carrying a specific fraction $x$ of the hadron's light-front momentum and squared four-momentum transfer $t$, and interpolate between collinear parton distribution functions (PDFs) in the forward limit and elastic form factors when integrated over $x$.
The latter include the Dirac and Pauli form factors, from the lowest moments of the GPDs, and gravitational form factors from the $x$-weighted GPD moments~\cite{Ji:1998pc} (for reviews 
see, {\it e.g.}, Refs.~\cite{Goeke:2001tz, Belitsky:2005qn}).
A further property of GPDs is that they obey different evolution equations in different kinematic regions: either the DGLAP (Dokshitzer-Gribov-Lipatov-Altarelli-Parisi) region as for collinear PDFs, where both the struck and returning partons carry positive momentum fractions of the initial proton momentum \cite{Altarelli:1977zs, Gribov:1972ri, Dokshitzer:1977sg}, or the ERBL (Efremov-Radyushkin-Brodsky-Lepage) region, which can be interpreted as describing a quark-antiquark pair emerging from the proton, as appropriate for distribution amplitudes (DAs)~\cite{Efremov:1979qk, Lepage:1980fj, Lepage:1979zb}.

On the experimental side, data from processes such as deeply-virtual Compton scattering (DVCS)~\cite{Ji:1996nm} and hard exclusive meson production (HEMP)~\cite{Radyushkin:1996ru, Mankiewicz:1997uy, Collins:1998be}, including from the H1~\cite{Adloff:2001cn, Aktas:2005ty, H1:2009wnw} and ZEUS~\cite{Breitweg:1998nh, Chekanov:2003ya} collaborations at HERA, the HERMES fixed target experiment~\cite{Airapetian:2001yk, Airapetian:2011uq, Airapetian:2012mq}, the COMPASS Collaboration at CERN~\cite{dHose:2004usi, Fuchey:2015frv}, and from the CLAS~\cite{Stepanyan:2001sm, CLAS:2012cna, Seder:2014cdc, Jo:2015ema, CLAS:2018ddh, CLAS:2021gwi} and Hall~A~\cite{Defurne:2015kxq, JeffersonLabHallA:2022pnx} collaborations at Jefferson Lab, have been used to provide indirect information on GPDs through the Compton form factors.
The latter are given as convolutions of GPDs with hard scattering kernels derived within QCD factorization~\cite{Collins:1998be, Belitsky:2001ns}. 
The extraction of GPDs and their moments has also been a motivation for planned experimental programs at future facilities such as the Electron-Ion Collider (EIC)~\cite{Accardi:2012qut, Avakian:2019csr}.

The extraction of GPDs directly from experimental data is of course the cornerstone of the quest to determine the 3-dimensional structure of the nucleon.
Pioneering studies have already been made by several groups worldwide on this effort~\cite{Diehl:2004cx, Kumericki:2009uq, Goldstein:2010gu, Guo:2022upw, Guo:2023ahv}, although direct, model-independent extractions remain a formidable challenge. 
Among some early studies, Diehl {\it et al.}~\cite{Diehl:2004cx} used a simple empirical parametrization of the $x$ and $t$ dependence of GPDs at zero skewness, with forward collinear PDFs as input.
Kumeri\v{c}ki and M\"{u}ller~\cite{Kumericki:2009uq} studied DVCS at small values of the Bjorken-$x$ variable, using conformal integral GPD moments, performing a first model-dependent extraction of the unpolarized GPD $H$ from HERA and Jefferson Lab DVCS data.
Goldstein {\it et al.}~\cite{Goldstein:2010gu} carried out a global analysis of DVCS observables, together with nucleon elastic form factors and deep-inelastic scattering measurements, using a flexible parametrization of GPDs inspired by a hybrid model of the nucleon as a quark-diquark system with Regge behavior. 
Most recently, Guo {\it et al.}~\cite{Guo:2022upw, Guo:2023ahv} employed the GUMP moment parametrization, motivated by the complex conformal spin partial wave expansion method of M\"{u}ller and Sch\"{a}fer \cite{Mueller:2005ed}, to perform a global analysis of GPDs from DVCS data, constrained by input on PDFs and elastic form factors, as well as recent lattice QCD calculations~\cite{Alexandrou:2020zbe, Alexandrou:2021jok}. 
Mamo and Zahed proposed a string-based parametrization using the Mellin-Barnes integral representations~\cite{Mamo:2024vjh}, which allows construction of the quark and gluon GPDs at any skewness.

Notwithstanding these important developments, in a seminal paper Bertone {\it et al.}~\cite{Bertone:2021yyz} recently pointed out, within a next-to-leading order QCD analysis, that a critical limitation of processes such as DVCS is their inability to uniquely determine the $x$ dependence of the GPDs due to the presence of so-called ``shadow GPDs.''
The shadow GPDs are a set of solutions to the inverse problem of extracting GPDs from DVCS data, which renders the extracted GPDs not unique.
Moffat {\it et al.}~\cite{Moffat:2023svr} further investigated the extent to which QCD evolution can provide constraints on the shadow GPDs, observing that given empirical data over a sufficiently wide range of skewness~$\xi$ and scale $Q^2$, the $Q^2$ evolution may in principle constrain the shadow GPDs, but over a rather limited range of $x$ and $\xi$.

More recently, Qiu and Yu~\cite{Qiu:2022pla, Qiu:2022bpq} explored a novel new set of exclusive processes, including single diffractive hard exclusive photoproduction, as a means of more directly constraining the $x$ dependence of GPDs.
Such processes can be factorized into process-independent GPDs and perturbatively calculable, infrared safe hard coefficients~\cite{Qiu:2022pla}. One can extract GPDs from polarized photoproduction cross sections, as well as asymmetries constructed from photon polarization and hadron spin, which could in future be measured, for example, at Jefferson Lab Hall~D.
Other related processes, such as exclusive production of high transverse momentum photons in $\pi N$ scattering, can be factorized into GPDs and pion DAs convoluted with short distance hard kernels \cite{Qiu:2022bpq}, if the photons' transverse momentum $q$ is $\gg \Lambda_{\rm QCD}$, with corrections suppressed by powers of $1/q$. The $x$ depdendence of GPDs can be also accessed through double DVCS~\cite{Belitsky:2002tf, Guidal:2002kt} by choosing different invariant masses of the produced lepton pair.
In addition, the photoproduction of a large mass diphoton $\gamma N \to \gamma\gamma N$ has been proposed to probe $C$-odd GPDs~\cite{Grocholski:2021man, Grocholski:2022rqj}.

Along with the empirical determinations of GPDs from data, developments in lattice QCD have allowed access to nucleon structures from first principles calculations, providing complementary information that is often difficult to obtain from experiment~\cite{Lin:2020rxa, Bhattacharya:2022aob, Alexandrou:2020zbe, Alexandrou:2006ru}.
In particular, considerable effort has been devoted to directly computing the 
$x$ dependence of distributions, using either the quasi-parton distribution~\cite{Ji:2013dva}, pseudo-distribution~\cite{Orginos:2017kos}, or lattice good cross sections~\cite{Ma:2014jla, Ma:2017pxb} approaches.

While the lattice continues to make progress with improved control over systematic uncertainties in the simulations, it is also useful to explore possible insights from model calculations, many of which have been performed over the past 25 years~\cite{Ji:1997gm, Petrov:1998kf, Penttinen:1999th, Goeke:2001tz, Choi:2001fc, Choi:2002ic, Tiburzi:2001je, Theussl:2002xp, Boffi:2002yy, Schweitzer:2002nm, Scopetta:2003et, Ossmann:2004bp, Wakamatsu:2005vk, Mineo:2005vs, Pasquini:2006ib, Goeke:2008jz}.
%
%
Even though their direct connection to QCD is not always transparent, they can nevertheless provide glimpses into some of the qualitative features and characteristics of GPDs.
%

In between the phenomenological models and lattice QCD simulations, constraints from chiral effective field theory (EFT) have been used to make predictions for various sea quark flavor asymmetries in the nucleon~\cite{Thomas:1983fh, Thomas:2000ny, Chen:2001nb, Salamu:2014pka, Luan:2023lmt}.
These are based on the observation that the long-range structure of the nucleon has contributions from the pseudoscalar meson cloud, associated with the model-independent leading nonanalytic behavior of observables expanded in a series of $m_\pi$.
Operationally, the sea quark distributions can be obtained through the convolution of a probability of the nucleon to split into a virtual meson and recoil baryon (``splitting function''), with the valence quark distribution of the meson.
The splitting function can be calculated from EFT, in which its moments are expanded as a series in the pseudoscalar meson mass ${\cal O}(m_\phi/\Lambda_{\chi})$ or small external momentum ${\cal O}(q/\Lambda_{\chi})$, where $\Lambda_{\chi}\sim$ 1~GeV is the 
scale associated with the chiral EFT.
Traditional EFT calculations based on dimensional regularization have found it challenging to describe lattice results in the large-$Q^2$ region, or with pion mass 400~MeV~\cite{Young:2002ib, Leinweber:2003dg}, since the short-distance effect arsing from the loop integral leads to a poor convergence of the chiral expansion~\cite{Donoghue:1998bs, Leinweber:2003dg}.
To improve the convergence, finite range regularization has been proposed and applied for the extrapolation of various quantities calculated by lattice QCD, including the vector meson mass, magnetic moments, magnetic and strange form factors, charge radii, and moments of PDFs and GPDs, from large pion masses to the physical mass~\cite{Young:2002ib, Leinweber:2003dg, Wang:2007iw, Allton:2005fb, Wang:1900ta, Wang:2012hj, Hall:2013dva, Shanahan:2012wh, Shanahan:2014uka}.
Here the three dimensional regulator is included in the loop integral and can be regarded as partially resumming the contribution from the higher order interaction.
In a similar vein, a {\it nonlocal} chiral effective Lagrangian has also been proposed~\cite{Wang:2010rib, He:2017viu, He:2018eyz}, in which the covariant regulator is automatically generated from the Lagrangian. 
One can obtain reasonable descriptions of the nucleon electromagnetic and strange form factors at large $Q^2$ using this method, which has also been applied to calculate the strange--antistrange PDF asymmetry $s-\bar{s}$~\cite{Salamu:2018cny, Salamu:2019dok}, the sea quark transverse momentum dependent (TMD) Sivers function~\cite{He:2019fzn}, and zero-skewness GPDs~\cite{He:2022leb}. 
More details on 
the nonlocal chiral effective theory can be found in Ref.~\cite{Wang:2022bxo}.
Recently, Copeland and Mehen~\cite{Copeland:2024wwm} also discussed a framework for matching TMD PDFs onto chiral effective theory operators in terms of TMD hadronic distribution functions calculated in chiral perturbation theory.

In this paper, we extend our previous analysis~\cite{He:2022leb} of zero-skewness GPDs in the proton within the nonlocal chiral effective theory to the nonzero skewness case. 
In Sec.~\ref{sec.theory} we present the basic theoretical framework used in our analysis, including the definition of unpolarized GPDs (Sec.~\ref{ssec.unpolgpd}) and the nonlocal meson-baryon interaction that underlies our calculation (Sec.~\ref{ssec.lagrangian}).
The one-loop nucleon splitting into meson plus octet and decuplet baryon splitting functions are given in Sec.~\ref{sec.splitting} from the set of rainbow and bubble diagrams that contribute to the antiquark GPDs in the nucleon.
The convolution formalism is derived in Sec.~\ref{sec.convolution}, where we outline the differences with the zero skewness GPD case. 
In particular, we separate the contribution to DGLAP and ERBL region, and prove that such a convolution formula can generate the skewness-independent form factors after integrating over $x$.
Numerical results for the two and three dimensions splitting functions and light antiquark GPDs using the convolution formulas are presented in Sec.~\ref{sec.numerical}.
Finally, Sec.~\ref{sec.summary} summarises our results and anticipates future extensions of this analysis.
In Appendix~\ref{sec.appendix}, we also demonstrate the GPDs' $n$-th moments obtained with convolution formula satisfy the general polynomial property.
In Appendix~\ref{sec.appendixB}, we summarize the explicit expressions for the splitting functions and the coupling constants presented in this study.

\section{Theoretical framework}
\label{sec.theory}

In this section we summarize the theoretical framework on which our analysis is based.
We begin with a summary of the definitions of the spin-averaged GPDs that are the focus of this analysis, followed by a discussion of the contributions to the GPDs from pseudoscalar meson loops, formulated within chiral effective theory.
We summarize the pertinent aspects of the chiral Lagrangian describing the meson-nucleon interaction, and present results for the nucleon to baryon $+$ meson splitting functions at nonzero skewness.

\subsection{Unpolarized GPDs}
\label{ssec.unpolgpd}

The spin-averaged GPDs for a quark of flavor $q$ in a nucleon are defined in terms of the Fourier transform of the nonforward matrix elements of quark bilocal field operators, taken between nucleon states with initial momentum $p$ and final momentum $p'$~\cite{Ji:1996nm},
\begin{eqnarray}
\label{eq:GPD}
\int_{-\infty}^{\infty}\frac{\dd{\lambda}}{2\pi} e^{-ix\lambda}
\langle p'| 
    \bar\psi_q (\tfrac12\lambda n) \slashed{n}\, \psi_q(-\tfrac12\lambda n)
|p \rangle 
&=& \bar u(p') 
  \Big[ \slashed n H_q(x,\xi,t) 
      + \frac{i\sigma^{\mu\nu}n_\mu \Delta_\nu}{2M}\, E_q(x,\xi,t)
  \Big] u(p),
\notag\\
& &
\end{eqnarray}
where the light-cone vector $n_\mu$ projects the ``plus'' component of momentum, $\lambda$ is a dimensionless parameter and $M$ is the nucleon mass.
Invariance under Lorentz transformations requires the Dirac ($H_q$) and Pauli ($E_q$) GPDs to be functions of the light-cone momentum fraction $x$ of the nucleon carried by the initial quark with momentum $k_q$,
\begin{equation}
x\,   \equiv\, \frac{k_q^+}{P^+},
\end{equation}
and the skewness parameter $\xi$,
\begin{equation}
\xi\, \equiv\, -\frac{\Delta^+}{2P^+},
\end{equation}
where
%
\begin{eqnarray}
P &=& \frac12 (p + p'), \qquad 
\Delta\, =\, p' - p
\end{eqnarray}
%
are the average and difference of the initial and final nucleon momenta, respectively.
We define the light-cone components of a four-vector $k^\mu$ by $k^\pm = \frac{1}{\sqrt2}(k^0 \pm k^3)$, and for convenience choose a symmetric frame of reference for the initial and final nucleon momenta,
\begin{subequations}
\begin{eqnarray}
p^\mu   &=& \Big(P^+-\frac{\Delta^+}{2},P^--\frac{\Delta^-}{2},-\frac{\bm{\Delta}_\perp}{2}\Big), \\
p'^\mu\ &=& \Big(P^++\frac{\Delta^+}{2},P^-+\frac{\Delta^-}{2},\quad \frac{\bm{\Delta}_\perp}{2}\Big).
\end{eqnarray}
\end{subequations}
In addition to being functions of $x$ and $\xi$, the GPDs also depend on the hadronic four-momentum transfer squared, $t \equiv \Delta^2$, as well as on the scale $Q^2$, which is typically taken to be the virtuality of the photon which couples to the hard scattering. The momentum transfer should satisfy the kinematic constraint $-\Delta^2 > 4 \xi^2 M^2/(1-\xi^2)$ in order to guarantee positive transferred momentum squared, $\bm{\Delta_\perp}^2 > 0$.

Integrating the GPDs $H_q$ and $E_q$ over the quark light-cone momentum fraction $x$, the Dirac and Pauli form factors for a quark flavor $q$ can be expressed as 
\begin{subequations}
\begin{eqnarray}
\label{eq.F1q}
\int_{-1}^1\,\dd{x} H_q(x,\xi,t) &=& F_1^q(t), \\
\label{eq.F2q}
\int_{-1}^1\,\dd{x} E_q(x,\xi,t) &=& F_2^q(t).
\end{eqnarray}
\end{subequations}%
Note that the $\xi$ dependence vanishes in the integrals of the GPDs over $x$.
Furthermore, the gravitational form factors $A^q$, $B^q$ and $C^q$ can be obtained from the first moments of the $H_q$ and $E_q$ GPDs as
\begin{subequations}
\begin{eqnarray}
\int_{-1}^1\,\dd{x} x\, H_q(x,\xi,t) &=& A^q(t) + (2\xi)^2\, C^q(t), \\
\int_{-1}^1\,\dd{x} x\, E_q(x,\xi,t) &=& B^q(t) - (2\xi)^2\, C^q(t),
\label{eq:Gffs}
\end{eqnarray}
\end{subequations}
where now explicit dependence on the skewness $\xi$ now appears on the right hand side.

\subsection{Chiral loop contributions to GPDs}
\label{ssec.lagrangian}

\begin{figure}[tp]
\begin{center}
\includegraphics[scale=1]{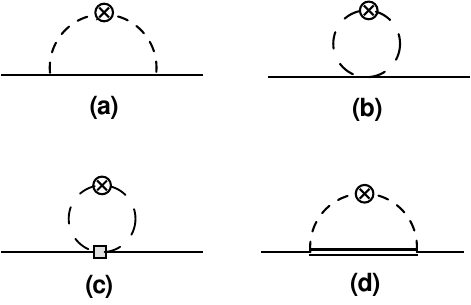}
\caption{One-loop diagrams for the proton to pseudoscalar meson (dashed lines) and octet baryon (solid lines) or decuplet baryon (double solid lines) splitting functions. The crossed circles ($\otimes$) represent the interaction with the external vector field from the minimal substitution, and the gray square (${\textcolor{gray}{\blacksquare}}$) denotes the last interaction term in Eq.~(\ref{eq:j1}). } 
\label{fig:diagrams}
\end{center}
\end{figure}

In our analysis we consider the contributions to the sea quark and antiquark GPDs in the proton arising from the virtual pseudoscalar meson cloud dressing of the bare baryon, as a first step towards a complete calculation to one-loop order.
Typically, in calculations of meson loop contributions to sea quark and antiquark asymmetries, assuming that the undressed proton has a flavor symmetric sea~\cite{Salamu:2014pka}, the meson coupling diagrams in Fig.~\ref{fig:diagrams} are the dominant source of differences between sea quark and antiquark PDFs and GPDs.

In our previous analysis~\cite{He:2022leb} we considered the nonforward nucleon to to virtual meson and recoil baryon splitting functions at zero skewness.
Generalizing to nonzero $\xi$, the splitting functions can be written as
\begin{equation}
\bar{u}(p')
\Big[ \gamma^+ f(y,\xi,t) + \frac{i\sigma^{+\nu}\Delta_\nu}{2M}\, g(y,\xi,t)
\Big] u(p)
= \int \dd[4]{k} \widetilde{\Gamma}^+(k)\, \delta\Big(y+\xi-\frac{k^+}{P^+}\Big)\,
\equiv\, \Gamma^+,
\label{eq.Gamma+def}
\end{equation}
where $f(y,\xi,t)$ and $g(y,\xi,t)$ are the corresponding Dirac- and Pauli-like splitting functions, and $y$ is the light-cone momentum fraction of the nucleon carried by the intermediate state hadron.
The operator $\widetilde{\Gamma}^\mu(k)$ is related to the matrix elements of the hadronic current $J^\mu$,
\begin{equation}
\label{eq.emvertex}
\langle N(p')|J^{\mu}|N(p) \rangle\,
=\, \bar{u}(p')
    \Big[ \gamma^\mu F_1^N(t) + \frac{i\sigma^{\mu\nu}\Delta_\nu}{2M}F_2^N(t)
    \Big]
    u(p)\,
\equiv
\int\!\dd[4]{k}\, \widetilde{\Gamma}^\mu(k),
\end{equation}
where $J^\mu$ is given in terms of its individual quark flavor contributions as in Ref.~\cite{He:2022leb}.
One can also verify that the Dirac and Pauli form factors can be obtained by integrating the splitting functions over $y$.

To compute the hadronic splitting functions $f$ and $g$, we use the standard chiral SU(2)$_L\times$SU(2)$_R$ effective Lagrangian, the details of which can be found, for example, in Refs.~\cite{Jenkins:1990jv, Bernard:2007zu, Hacker:2005fh}. 
In the one-loop approximation used in our analysis, the chiral Lagrangian 
can be expanded to reveal the interaction terms, given by
\begin{eqnarray}
{\cal L}_{\rm int}
&=& \frac{g_A}{2f_\pi}
  \left( \bar p\, \gamma^\mu \gamma_5 p\, \partial_\mu \pi^0
       + \sqrt2\, \bar p\, \gamma^\mu \gamma_5 n\, \partial_\mu \pi^+
  \right)  
\notag\\
&+& \frac{\cal C}{\sqrt{12} f_\pi}
  \left(
  - 2\, \bar p\, \Theta^{\nu\mu} \Delta_\mu^+\, \partial_\nu \pi^0
  - \sqrt2\, \bar p\, \Theta^{\nu\mu} \Delta_\mu^0\, \partial_\nu \pi^+
  + \sqrt6\, \bar p\, \Theta^{\nu\mu} \Delta_\mu^{++}\, \partial_\nu \pi^-
  \right)
\notag\\
&+& \frac{i}{4 f_\pi^2}\, \bar p\, \gamma^\mu p\,
    (\pi^+ \partial_\mu \pi^-  -  \pi^- \partial_\mu \pi^+)
+ \frac{2i (b_{10}+b_{11})}{f_\pi^2}\, \bar p\, \sigma^{\mu\nu} p\,   
    \partial_\mu\pi^+\, \partial_\nu \pi^-  + {\rm H.c.},
\label{eq:j1}
\end{eqnarray}
where H.c. refers to the Hermitian conjugate, and $f_\pi=92.4$~MeV is the pion decay constant.
The axial vector coupling constant for the octet baryon is set to the standard value $g_A = 1.26$, and the octet-decuplet transition coupling ${\cal C}$ is chosen to be $-\frac65 g_A$ from SU(6) spin-flavor symmetry.
The octet–decuplet transition operator $\Theta^{\mu\nu}$ is defined as~\cite{Nath:1971wp}
\begin{equation}
\Theta^{\mu\nu}
= g^{\mu\nu} - \gamma^\mu \gamma^\nu.
\label{eq:Theta}
\end{equation}
The coefficients $b_{10}$ and $b_{11}$ in Eq.~(\ref{eq:j1}) for the nucleon-pion contact interaction have previously been determined to be
    $b_{10}=1.24$~GeV$^{-1}$,
and
    $b_{11}=0.46$~GeV$^{-1}$~\cite{Kubis:2000aa}.
In our previous work~\cite{He:2022leb} we discussed in detail the construction of the nonlocal chiral Lagrangian; here we simply present the final nonlocal interaction ${\cal L}^{\rm (nonloc)}_{\rm int}$ we used in this calculation,
\begin{eqnarray}
{\cal L}^{\rm (nonloc)}_{\rm int}(x)
&=& \bar{p}(x)
    \bigg( \frac{g_A}{2f_\pi}\, \gamma^\mu \gamma_5 p(x)
	 - \frac{\cal C}{\sqrt{3} f_\pi}\, \Theta^{\mu\nu} \Delta^+_\nu(x)
    \bigg)
    \!\int\!\dd[4]{a}\, F(a)\, \partial_\mu \pi^0(x+a)
\notag\\
& & \hspace*{-2.5cm}
+\ \bar{p}(x)
    \bigg( \frac{g_A}{\sqrt{2}f_\pi}\, \gamma^\mu \gamma_5 n(x)
	 - \frac{\cal C}{\sqrt{6} f_\pi}\, \Theta^{\mu\nu} \Delta^0_\nu(x)
    \bigg)
    \!\int\!\dd[4]{a}\, F(a)\, \partial_\mu \pi^+(x+a)
\notag\\
& & \hspace*{-2.5cm}
+\ \frac{\cal C}{\sqrt{2} f_\pi}\,\bar{p}(x)\Theta^{\mu\nu} \Delta^{++}_\nu(x)
    \!\int\!\dd[4]{a}\, F(a)\, \partial_\mu \pi^-(x+a)
	+ {\rm H.c.}
\notag\\
& & \hspace*{-2.5cm}
 +\ \frac{i}{4f_\pi^2}\,
    \bar{p}(x) \gamma^\mu p(x)
    \int\!\dd[4]{a}\!\int\!\dd[4]{b}\ F(a)\, F(b)\,
    \left[ \pi^+(x+a)\, \partial_\mu \pi^-(x+b)
	 - \partial_\mu \pi^+(x+a) \pi^-(x+b)
    \right]
\notag\\
& & \hspace*{-2.5cm}
 +\ \frac{2i (b_{10}+b_{11})}{f_\pi^2}\,
    \bar{p}(x) \sigma^{\mu\nu} p(x)
    \int\!\dd[4]{a}\!\int\!\dd[4]{b}\ F(a)\, F(b)\,
    \partial_\mu\pi^+(x+a)\, \partial_\nu \pi^-(x+b).
\label{eq:Lnonloc_had}
\end{eqnarray}
The function $F(a)$ here describes the strength of the correlation between the baryon and the meson, and will become the regulator function in the momentum space.
In the limit when $F(a) \to \delta(a)$, one can show that Eq.~(\ref{eq:Lnonloc_had}) reduces to the local version of the Lagrangian given in Eq.~(\ref{eq:j1}).

\subsection{Nonzero skewness splitting functions}
\label{sec.splitting}

In this work we will be particularly concerned with the contributions to proton GPDs from the pseudoscalar meson coupling diagrams in Fig.~\ref{fig:diagrams}, as a first step towards a complete one-loop calculation of the GPDs at nonzero skewness.
Contributions to GPDs from addition couplings to the octet and decuplet baryons, including baryon rainbow and Kroll-Ruderman diagrams~\cite{He:2022leb}, will be investigated in future work.

The contribution to the nonzero skewness splitting functions from the rainbow diagram in Fig.~\ref{fig:diagrams}(a) with the pseudoscalar meson and octet baryon is given by
\begin{eqnarray} 
\bar{u}(p')\, \Gamma_{(\rm a)}^+\, u(p) 
&=& \bar{u}(p')\,
\frac{C^2_{B\phi}}{f^2}
\int\!\frac{\dd[4]{k}}{(2\pi)^4}
(\slashed{k}+\slashed{\Delta})\gamma_5\,
\widetilde{F}(k+\Delta) 
\frac{i}{D_\phi(k+\Delta)}\, (2k^+ + \Delta^+)
\nonumber\\
&&\hspace*{2.3cm}
\times \frac{i}{D_\phi(k)} \frac{i(\slashed{p}-\slashed{k}+M_B)}{D_B(p-k)}
\gamma_5 \slashed{k}\, 
\widetilde{F}(k)\,
\delta\Big(y+\xi-\frac{k^+}{P^+}\Big)\,
u(p)                                          \nonumber  \\
&\equiv& \bar{u}(p')
\bigg[
  \gamma^+ f_{\phi B}^{({\rm rbw})}(y,\xi,t)
  + \frac{i\sigma^{+\nu}\Delta_\nu}{2M} g_{\phi B}^{({\rm rbw})}(y,\xi,t)
\bigg] u(p),                       
\label{eq.splitfnproj}
\end{eqnarray}
where the propagator factors $D_\phi(k)$ and $D_B(p)$ are defined as
\begin{equation} 
D_\phi(k) = k^2 - m_\phi^2 + i\epsilon,
\qquad\qquad
D_B(p)    = p^2 - M_B^2 + i\epsilon,
\end{equation}
and $m_\phi$ and $M_B$ represent the pseudoscalar meson and baryon masses, respectively.
The function $\widetilde{F}$ is a regulator function, obtained by performing a Fourier transformation on the factor $F(a)$ in Eq.~(\ref{eq:Lnonloc_had}), and is used to regularize the ultraviolet divergence in the loop integration.
For simplicity we choose this to be a function of the meson momentum only (see Sec.~\ref{sec.numerical} below). 
The explicit expressions for the splitting functions $f_{\phi B}^{({\rm rbw})}$ and $g_{\phi B}^{({\rm rbw})}$ and the couplings $C^2_{B\phi}$ are given in Appendix~\ref{sec.appendixB}.

The bubble diagrams are illustrated in Figs.~\ref{fig:diagrams}(b) and \ref{fig:diagrams}(c).
For the regular bubble diagram in Fig.~\ref{fig:diagrams}(b), the corresponding splitting function is given by
\begin{eqnarray} 
\bar{u}(p')\, \Gamma_{(\rm b)}^+\, u(p) 
&=&-\bar{u}(p')\,
\frac{C_{\phi\phi}}{2f^2}
\int\!\frac{\dd[4]{k}}{(2\pi)^4}\,
i(2\slashed{k}+\slashed{\Delta})\,
\widetilde{F}(k+\Delta)
\widetilde{F}(k)\,
\frac{i}{D_\phi(k+\Delta)}
(2k^++\Delta^+)
\notag\\
&&\hspace*{2.3cm}
\times
\frac{i}{D_\phi(k)}\,
\delta\Big(y+\xi-\frac{k^+}{P^+}\Big)\,
u(p),
\notag\\
&\equiv& \bar{u}(p') \gamma^+ u(p)\,
f_{\phi}^{\rm (bub)}(y,\xi,t).
\label{eq.f_phi_bub}
\end{eqnarray}
The additional bubble diagram in Fig.~\ref{fig:diagrams}(c) is derived from the last interaction term in Eq.~(\ref{eq:j1}), and its contribution to the matrix element can be written as
\begin{eqnarray} 
\bar{u}(p')\, \Gamma_{(\rm c)}^+\, u(p) 
&=& -\bar{u}(p')\,
\frac{C'_{\phi\phi}}{f^2}
\int\!\frac{\dd[4]{k}}{(2\pi)^4}\,
\sigma^{\lambda\nu}\Delta_\lambda k_\nu\, \widetilde{F}(k+\Delta)
\widetilde{F}(k)
\frac{i}{D_\phi(k+\Delta)}
(2k^++\Delta^+)
\notag\\
&&\hspace*{2.3cm}
\times
\frac{i}{D_\phi(k)}\,
\delta\Big(y+\xi-\frac{k^+}{P^+}\Big)\,
u(p),
\notag\\
&\equiv& \bar{u}(p') \gamma^+ u(p)\
g_{\phi}^{\prime \rm (bub)}(y,\xi,t),
\label{eq.f_phi_bub_add}
\end{eqnarray}
following the convention of Ref.~\cite{He:2022leb}.

Finally, for the decuplet baryon rainbow diagram in Fig.~\ref{fig:diagrams}(d), the contribution to $\Gamma^+$ is given by a form similar to that for the octet baryon diagram in Fig.~\ref{fig:diagrams}(a),
\begin{eqnarray} 
\bar{u}(p')\, \Gamma_{(\rm d)}^+\, u(p) 
&=&-\bar{u}(p')\,
\frac{C_{T\phi}^2}{f^2}
\int\!\frac{\dd[4]{k}}{(2\pi)^4}\,
(k+\Delta)_\lambda\,
\Theta^{\lambda\alpha}
\widetilde{F}(k+\Delta)\frac{i}{D_\phi(k+\Delta)}\, (2k^++\Delta^+)
\notag\\
&&\times
\frac{i}{D_\phi(k)}
\frac{i}{\slashed{p}-\slashed{k}-M_T}
S_{\alpha\beta}(p-k)\, \Theta^{\beta\rho} k_\rho\,
\widetilde{F}(k)\,
\delta\Big(y+\xi-\frac{k^+}{P^+}\Big)\,
u(p)
\notag\\
&\equiv&
\bar{u}(p')
\bigg[
  \gamma^+ f_{\phi T}^{({\rm rbw})}(y,\xi,t) 
+ \frac{i\sigma^{+\nu}\Delta_\nu}{2M}g_{\phi T}^{({\rm rbw})}(y,\xi,t)
\bigg]
u(p),
\end{eqnarray}
where $M_T$ is the mass of baryon decuplet.
Again, the expressions for the integrals of the splitting functions $f_{\phi T}^{({\rm rbw})}$ and $g_{\phi T}^{({\rm rbw})}$, along with the coefficients $C^2_{T\phi}$, as well as the corresponding bubble distributions $f_{\phi}^{\rm (bub)}$ and $g_{\phi}^{\prime \rm (bub)}$, are given in Appendix~\ref{sec.appendixB}.

\section{GPDs from pseudoscalar meson loops}
\label{sec.convolution}

The recent analysis in Ref.~\cite{He:2022leb} derived formulas for quark and antiquark GPDs at zero skewness in terms of convolutions of the nucleon to pseudoscalar meson plus baryon splitting functions, and GPDs associated with the mesons and baryons in the intermediate state.
In this section, we discuss the extension of this formalism to the case of nonzero skewness.
This exercise requires paying careful attention to the different regions of parton momentum fraction $x$, meson light-cone momentum fraction $y$, and skewness parameter $\xi$.

\begin{figure}[tp]
\begin{center}
\includegraphics[scale=1]{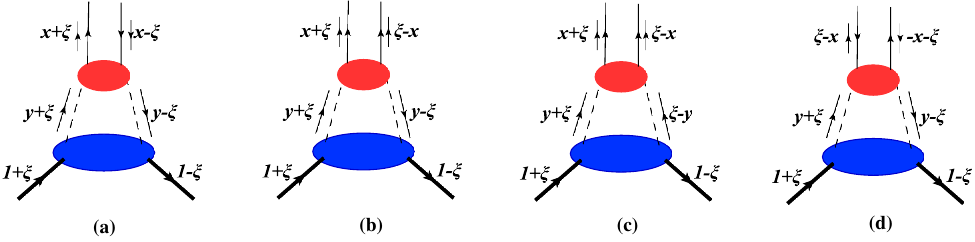}
\caption{Representation of the convolution formula in Eqs.~(\ref{eq:conv_a}), (\ref{eq:conv_b}), (\ref{eq:conv_c}), and (\ref{eq:conv_d}), respectively, with the \{dashed, thick solid, thin solid\} lines representing the \{pseudoscalar meson, proton, quark\}. The processes in diagrams (a) and (d) represent the DGLAP region for the quark and antiquark, respectively, while the processes in (b) and (c) contribute to the ERBL region.}
\label{fig:convol_illu}
\end{center}
\end{figure}

For the rainbow diagrams in Figs.~\ref{fig:diagrams}(a) and \ref{fig:diagrams}(d), the relevant splitting functions are nonzero when $y$ is in the region $-\xi \leq y \leq 1$, for positive $\xi$ values. 
The splitting function in this region can be convoluted with the pion GPD to obtain the quark GPD in the 
nucleon. 
Figure~\ref{fig:convol_illu} illustrates the decomposition of the convolution formula for the rainbow diagram in Fig.~\ref{fig:diagrams}(a) into different subprocesses.
The splitting functions for Figs.~\ref{fig:convol_illu}(a), \ref{fig:convol_illu}(b) and \ref{fig:convol_illu}(d) are all located in the $y>\xi$ region, while the quark GPD in the pion belongs to the quark DGLAP, ERBL and antiquark DGLAP regions, respectively.
After convoluting with the hadronic splitting function, these contribute to the quark GPD of the proton in the quark DGLAP, ERBL, and antiquark DGLAP regions, respectively. 
For the subprocess in Fig.~\ref{fig:convol_illu}(c) the splitting function is located at $y\in [-\xi,\xi]$, and can be considered as the meson--meson pair annihilation process, analogous to the DA-like dynamics at the quark level.

By combining these various processes, the contribution to the skewness nonzero GPDs from the diagram in Fig.~\ref{fig:diagrams}(a) can be expressed in the convolution form as
\begin{small}
\begin{subequations}
\label{eq:convolution_sum}
\begin{empheq}[left={\hspace*{-0.4cm}
H_q^{\rm (rbw)}(x,\xi,t)=\empheqlbrace}]{align}
\label{eq:conv_a}
    &\int_x^1 \frac{\dd{y}}{y} f_{\phi B}^{({\rm rbw})}(y,\xi,t)\, H_{q/\phi}\Big(\frac{x}{y},\frac{\xi}{y},t\Big), \hspace*{4.6cm}    
    [\xi < x < y]
\\ \nonumber \\ 
\label{eq:conv_b}
    &\int_\xi^1 \frac{\dd{y}}{y} f_{\phi B}^{({\rm rbw})}(y,\xi,t)\, H_{q/\phi}\Big(\frac{x}{y},\frac{\xi}{y},t\Big), \hspace*{4.6cm}    
    [x < \xi < y]
\\ \nonumber \\
\label{eq:conv_c}
	&\int_{-\xi}^{\xi} \frac{\dd{y}}{2y}\, f_{\phi B}^{({\rm rbw})}(y,\xi,t)\frac{1}{\pi} \int_{s_0}^\infty\!\! \dd{s}
    \frac{{\rm Im}\Phi_{q/\phi}{ \big( \frac12(1\!+\!\frac{x}{\xi}), 
                                       \frac12(1\!+\!\frac{y}{\xi}),
                                      s
                                 \big)}}{s-t+i\epsilon},  \hspace*{0.52cm}    
    [|x|,|y| < \xi]\!\!
\\ \nonumber \\
\label{eq:conv_d}
    &\int_{-x}^1 \frac{\dd{y}}{y} f_{\phi B}^{({\rm rbw})}(y,\xi,t)\, H_{q/\phi}\Big(\frac{x}{y},\frac{\xi}{y},t\Big), \hspace*{3.6cm}    
    [\xi < -x < y < 1]
\end{empheq}
\end{subequations}
\end{small}%
where $H_{q/\phi}$ and $\Phi_{q/\phi}$ represent the valence GPD and generalized distribution amplitude (GDA), respectively, in the intermediate pseudoscalar meson.
These can be expressed, respectively, as~\cite{Diehl:2003ny}
\begin{subequations}
\begin{eqnarray}
 H_{q/\phi}(x,\xi,t) &=& 
\int_{-\infty}^{\infty}\frac{\dd{\lambda}}{4\pi} e^{-ix\lambda}\langle \phi(p')| 
    \bar\psi_q (\tfrac12\lambda n) \slashed{n}\, \psi_q(-\tfrac12\lambda n)
|\phi(p) \rangle,   \\
 \Phi_{q/\phi}(\eta,z,t) &=& 
\int_{-\infty}^{\infty}\frac{\dd{\lambda}}{2\pi} e^{-i(2\eta-1)\lambda}\langle 0| 
    \bar\psi_q (\tfrac12\lambda n) \slashed{n}\, \psi_q(-\tfrac12\lambda n)
|\phi(p')\phi(p) \rangle,
\end{eqnarray}
\end{subequations}
where $z=p'^+ / (p^+ + p'^+)$
and $f_{\phi B}^{({\rm rbw})}$ is the rainbow splitting function defined in Eq.~(\ref{eq.splitfnproj}).
The corresponding convolution formula for the quark magnetic GPD 
$E_q(x,\xi,t)$ can be obtained by replacing $f_{\phi B}^{({\rm rbw})}(y,\xi,t)$ by the Pauli splitting function $g_{\phi B}^{({\rm rbw})}(y,\xi,t)$. 
Similarly, the contribution from the pseudoscalar meson--decuplet diagram in Fig.~\ref{fig:diagrams}(d) can be obtained by replacing the octet splitting functions $\{f,g\}_{\phi B}^{(\text{rbw})}(y,\xi,t)$ by the corresponding decuplet functions $\{f,g\}_{\phi T}^{({\rm rbw})}(y,\xi,t)$.

The splitting functions appearing in Eqs.~(\ref{eq:conv_a}), (\ref{eq:conv_b}) and (\ref{eq:conv_d}) are located in the region $\xi < y$, and represent the contribution to the quark GPD in the DGLAP region, the quark GPD in the ERBL region, and the antiquark GPD in the DGLAP region, respectively.
The convolution in Eq.~(\ref{eq:conv_c}) also contributes to the GPD in the ERBL region from the splitting function at $y < |\xi|$, with the splitting function convoluted with the GDA of the meson, defined in the timelike region. 
The results in the spacelike region can be obtained by analytically continuing the intergral over $s$ in Eq.~(\ref{eq:conv_c}), where $s_0\sim 2m_\phi$ is the threshold for the meson pair production.
The total contribution of the rainbow diagrams in Figs.~\ref{fig:diagrams}(a) and \ref{fig:diagrams}(d) to the GPDs are then obtained by summing the four subprocesses in Fig.~\ref{fig:convol_illu}.

For the bubble diagrams in Figs.~\ref{fig:diagrams}(b) and \ref{fig:diagrams}(c), the splitting function is nonzero only when the meson momentum fraction $y$ is in the range $y \in [-\xi,\xi]$, so that these only contribute to GPDs in the ERBL region.
Their contributions to the quark electric GPD in the proton can be written as
\begin{eqnarray}
\label{eq:convolution_bub}
H_q^{(\rm bub)}(x,\xi,t)&=
	&\int_{-\xi}^{\xi} \frac{\dd{y}}{2y}\, f_{\phi}^{(\rm bub)}(y,\xi,t)\,
    \frac{1}{\pi} \int_{s_0}^\infty \dd{s}\frac{ \text{Im}
    \Phi_{q/\phi}{ 
    \big( 
    \frac12(1\!+\!\frac{x}{\xi}), \frac12(1\!+\!\frac{y}{\xi}), s
    \big)}}{s-t+i\epsilon},
\end{eqnarray}
where the splitting function for the bubble diagram, $f_{\phi}^{(\rm bub)}$, is given in Eq.~(\ref{eq.f_phi_bub}).
The analogous expression for the magnetic GPD 
$E_q^{(\rm bub)}$
is obtained from Eq.~(\ref{eq:convolution_bub}) by replacing the bubble splitting function $f_{\phi}^{(\rm bub)}$ by the corresponding function $g_\phi^{\prime \rm (bub)}$ introduced in Eq.~(\ref{eq.f_phi_bub_add}).

Integrating the GPDs over $x$ from $-1$ to $1$, we can write the lowest moment of $H_q^{(\rm rbw)}$ in terms of the pseudoscalar meson elastic form factor.
Adding the contributions from Eqs.~(\ref{eq:conv_a}), (\ref{eq:conv_b}) and (\ref{eq:conv_d}), integrated over the intervals $[\xi,1]$, $[-\xi,\xi]$ and $[-1,-\xi]$, respectively, we have
\begin{eqnarray}
\label{eq:xidepen1}
&&
\!\int_\xi^1 \!\dd{y} f_{\phi B}^{(\rm rbw)}(y,\xi,t)
\Bigg[
  \int_{\xi/y}^1\! \dd{z} H_{q/\phi}\Big(z,\frac{\xi}{y},t\Big)
+ \int_{-1}^{-\xi/y}\!\!\! \dd{z} H_{q/\phi}\Big(z,\frac{\xi}{y},t\Big)
+ \int_{-\xi/y}^{\xi/y}\!\! \dd{z} H_{q/\phi}\Big(z,\frac{\xi}{y},t\Big)
\Bigg]
\nonumber\\
&& \hspace*{3.5cm}
= \int_\xi^1 \dd{y} f_{\phi B}^{(\rm rbw)}(y,\xi,t)\, F_\phi(t),
\end{eqnarray}
where we have used the relation between the lowest moment of the quark GPD in the pseudoscalar meson and the meson form factor.
For the ERBL region in Eq.~(\ref{eq:conv_c}), the contribution to the integral of the GPD can be written as 
\begin{eqnarray} 
\label{eq:xidepen2}
\int_{-\xi}^\xi \dd{x} H_q^{(\rm rbw)}(x,\xi,t)
&=& \int_{-\xi}^\xi \dd{y}\, 
f_{\phi B}^{(\rm rbw)}(y,\xi,t) 
\frac{\xi}{\pi y} \int_{-1}^1 
\frac{\dd{z}}{2} \int_{s_0}^\infty \dd{s} 
\frac{\text{Im} \Phi_{q/\phi} 
\big( \frac12 (1+z), \frac12 (1+\frac{y}{\xi}), s \big)}{s-t+i\epsilon}    
\nonumber\\
&=&\int_{-\xi}^\xi \dd{y}\, f_{\phi B}^{(\rm rbw)}(y,\xi,t) 
\frac{\xi}{\pi y} \int_{s_0}^\infty \dd{s} 
\frac{\text{Im} \int_0^1 \dd{\eta}\, \Phi_{q/\phi}({\eta,\frac12 (1+\frac{y}{\xi}), s)}}{s-t+i\epsilon}  
\nonumber\\
&=&\int_{-\xi}^\xi \dd{y}\,
f_{\phi B}^{(\rm rbw)}(y,\xi,t) \frac{1}{\pi} \int_{s_0}^\infty \dd{s} \frac{\text{Im}F_\phi(s)}{s-t+i\epsilon}     
\nonumber\\ \hspace*{3.5cm}
&=&\int_{-\xi}^\xi \dd{y}\, f_{\phi B}^{(\rm rbw)}(y,\xi,t)\, F_\phi(t),
\end{eqnarray}
where we have used the relation 
$\int_0^1 \dd{\eta} \Phi_{q/\phi}(\eta,\kappa,s) = (2\kappa-1) F_\phi(s)$. 
Combining Eqs.~(\ref{eq:xidepen1}) and (\ref{eq:xidepen2}), one then obtains the lowest moment of the GPD, 
\begin{eqnarray} 
\int_{-\xi}^1 \dd{x}\, H_q^{(\rm rbw)}(x,\xi,t) 
&=& \int_{-\xi}^1 \dd{y}\, f_{\phi B}^{(\rm rbw)}(y,\xi,t)\, F_\phi(t), 
\end{eqnarray}
which corresponds to the Dirac form factor according to Eqs.~(\ref{eq.Gamma+def}) and (\ref{eq.emvertex}).

In the numerical calculations in this analysis, we consider specifically the case of the pion, $\phi = \pi$.
For the pion GPDs we use the Radyushkin's double distribution 
parametrization from Refs.~\cite{Radyushkin:1998bz, Musatov:1999xp}, which was also utilized by Amrath {\it et al.}~\cite{Amrath:2008vx},
\begin{eqnarray}
\label{eq:inputGPD} 
H_{q/\pi}(x,\xi,t)
&=& \int_{-1}^1 \dd{\beta} \int_{-1+|\beta|}^{1-|\beta|} \dd{\alpha}\, \delta(x-\beta-\xi \alpha)\, h_b(\beta,\alpha)\, H_{q/\pi}(\beta,0,t)  
\nonumber\\
&+&\frac{\xi}{|\xi|} D_{q/\pi}\Big(\frac{x}{\xi},t\Big)\, \theta(\xi-|x|),
\end{eqnarray}
where the function $h_b$ is defined by
\begin{eqnarray} 
h_b(\beta,\alpha) = \frac{\Gamma(2b+2)}{2^{2b+1}\Gamma^2(b+1)}
\frac{\big[(1-|\beta|)^2-\alpha^2\big]^b}{(1-|\beta|)^{2b+1}}.
\end{eqnarray}
We take the parameter $b=2$, and the GPD in the zero skewness case is defined according to the {\it ansatz}, $H_{q/\pi}(\beta,0,t) = H_{q/\pi}(\beta,0,0)\, F_\pi(t)$. 
The second term in Eq.~(\ref{eq:inputGPD}) represents the contribution to the $D$-term of the gravitational form factor of the pion, which is related to the pressure distribution of partons 
in the hadron.
To obtain the correct first moment, this term must be an odd function of $x$, and in the present analysis we use 
the simple form,
\begin{eqnarray}\label{eq:Dpit}
D_{q/\pi}(z,t)=\frac{15}{4} z(1-z^2)\, D_{q/\pi}(t),   
\end{eqnarray}
where $D_{q/\pi}(t)$ is the gravitational form factor of the pion, for which we choose a monopole form, $D_{q/\pi}(t) = D_{q/\pi}(0) / (1-t/\Lambda_\pi^2)$, with normalization $D_{q/\pi}(0) = -0.157$ and cutoff paramaeter $\Lambda_\pi=1.44$~GeV from the recent lattice QCD calculation by Hackett {\it et al.}~\cite{Hackett:2023nkr}.

In the forward limit the quark GPD in the pion can be related to the quark PDF, 
    $H_{q/\pi}(x,0,0) = q_\pi(x)$ for $x>0$, and 
    $H_{q/\pi}(x,0,0) = -\bar{q}_\pi(-x)$ for $x<0$
by crossing symmetry.
Since we only consider the lowest Fock state of the pion, composed of a valence quark and antiquark, we set 
    $q_\pi(x) = q^v_\pi(x)$,
the valence quark PDF in the pion, and 
    $\bar{q}_\pi(x) = 0$
for the sea quarks in the pion.
Recent analyses of pion PDFs have been performed by several groups~\cite{Barry:2018ort, Novikov:2020snp, Barry:2021osv, Kotz:2023pbu}; for simplicity we use the older parametrization from Ref.~\cite{Aicher:2010cb}, which is 
given at the 
scale $\mu=0.63$ GeV.

For the contribution to the pion GPD from the ERBL region, we need in addition information about the pion GDA, $\Phi_{q/\phi}$ [Eq.~(\ref{eq:conv_c})].
For this we also use the double parametrization of the form~\cite{Teryaev:2001qm, Kivel:2002ia, Diehl:2003ny}
\begin{eqnarray} 
\Phi_{q/\pi}(\eta,z,s)
&=& 2 (2z-1) \int_{-1}^1 \dd{\beta} \int_{-1+|\beta|}^{1-|\beta|}\!\! \dd{\alpha} \delta\big( (2\eta-1)-(2z-1)\beta-\alpha)\big)\,
h_b(\beta,\alpha)\, H_{q/\pi}(\beta,0,s)
\nonumber\\
& & 
\nonumber\\
&+& 2 D_{q/\pi}(2\eta-1,s).
\end{eqnarray}
Such a parametrization can be shown to satisfy the polynomial property in Eq.~(\ref{eq:GDA_pro}).

As discussed above, in this calculation we only consider contributions to the GPDs from direct couplings to the virtual pion loop, as in Fig.~\ref{fig:diagrams}.
Using quark flavor symmetry, the pion GPDs and pion GDAs for the different possible charge states can be related according to
\begin{subequations}
\begin{eqnarray} 
H_{u/\pi^+}
&=& H_{\bar{d}/\pi^+}
 =  H_{d/\pi^-}
 =  H_{\bar{u}/\pi^-}
 = 2\, H_{q/\pi^0}, \\
%
%
\Phi_{u/\pi^+}
&=& \Phi_{\bar{d}/\pi^+}
 =  \Phi_{d/\pi^-}
 =  \Phi_{\bar{u}/\pi^-}
 = 2\, \Phi_{q/\pi^0},
\end{eqnarray}
\end{subequations}
respectively, where $q = u, \bar u, d$ or $\bar d$ for the quark flavors in the neutral $\pi^0$.
Combining the contributions from the different diagrams in Fig.~\ref{fig:diagrams}, including rainbow and bubble, and the different isospin channels, the $u$-quark GPD in the proton can be written as
\begin{eqnarray}
H_u(x,\xi,t)
&=& H^{({\rm rbw})\pi^+ n}_u(x,\xi,t)
 +  H^{({\rm rbw})\pi^+ \Delta^0}_u(x,\xi,t)
 +  H^{({\rm rbw})\pi^0 p}_u(x,\xi,t)
 +  H^{({\rm rbw})\pi^0 \Delta^+}_u(x,\xi,t)
\nonumber\\
&+& H^{({\rm bub})\pi^+\pi^-}_u(x,\xi,t)
\nonumber\\
&-& H^{({\rm rbw})\pi^0 p}_{\bar{u}}(-x,\xi,t)
 -  H^{({\rm rbw})\pi^0\Delta^+}_{\bar{u}}(-x,\xi,t)
 -  H^{({\rm rbw})\pi^-\Delta^{++}}_{\bar{u}}(-x,\xi,t),
\label{eq.Hu_alldiags}
\end{eqnarray}
where we have explicitly written out the specific pion-baryon charge contributions, and similarly for the $u$-quark magnetic GPD $E_u$.
The results of $d$-quark GPDs can be obtained using the isospin symmetry 
$\{H,E\}_d(x,\xi,t)=-\{H,E\}_u(-x,\xi,t)$,
since we only consider the pion loop contribution in this work.
Note that Eq.~(\ref{eq.Hu_alldiags}) includes contributions from the quark DGLAP, ERBL, and antiquark DGLAP regions.
As mentioned above, contributions from the baryon coupling rainbow diagrams, as well as Kroll-Ruderman diagrams~\cite{He:2022leb}, will be investigated in future work~\cite{Gao:future}.

\section{Phenomenology of meson loop contributions to GPDs}
\label{sec.numerical}

In this section we will present the phenomenological results for the generalized splitting functions derived in the preceeding sections, along with the meson loop contributions to the light quark GPDs obtained using the convolution formula in Eq.~(\ref{eq:convolution_sum}).
We also apply the formalism to calculation of the Dirac, Pauli, and gravitational form factors, obtained from the lowest two moments of the GPDs. 
In the numerical calculations the loop integrals are regularized using a covariant regulator of a dipole form,
\begin{equation}
\widetilde{F}(k) = \left(\frac{\Lambda^2-m^2_\phi}{\Lambda^2-k^2}\right)^2,
\end{equation} 
with $\Lambda$ a mass parameter.
In practice, we use the value $\Lambda = 1.0(1)$~GeV obtained from previous analyses of meson loop contributions to PDFs and GPDs~\cite{He:2017viu, He:2018eyz, He:2022leb}.

\subsection{Splitting functions}
\label{sec.numerical_sf}

\begin{figure}[htbp]
\begin{minipage}[b]{.45\linewidth}
\hspace*{-0.3cm}\includegraphics[width=1.1\textwidth, height=5.5cm]{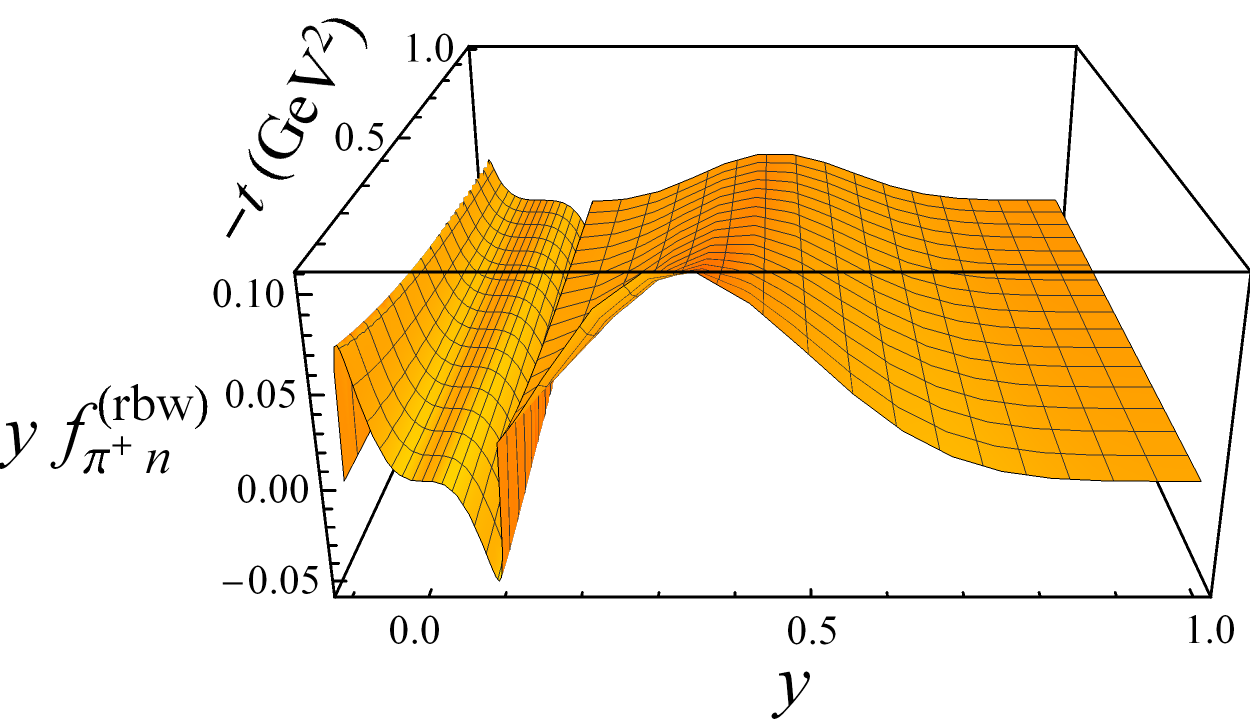}   
 \vspace{0pt}
\end{minipage}
\hfill
\begin{minipage}[b]{.45\linewidth}   
\hspace*{-0.95cm} \includegraphics[width=1.1\textwidth, height=5.5cm]{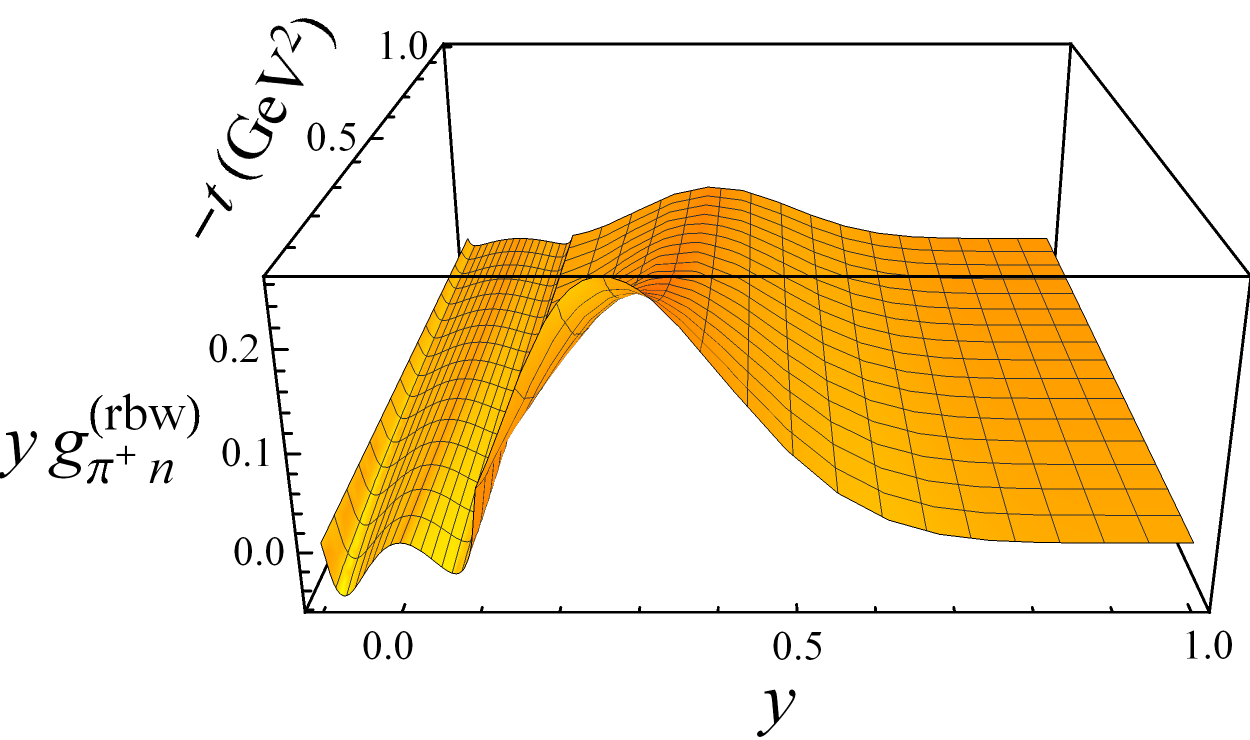} 
   \vspace{0pt}
\end{minipage}  
\\[-0.4cm]
\begin{minipage}[t]{.45\linewidth}
\hspace*{-0.5cm}\includegraphics[width=1.1\textwidth, height=5.5cm]{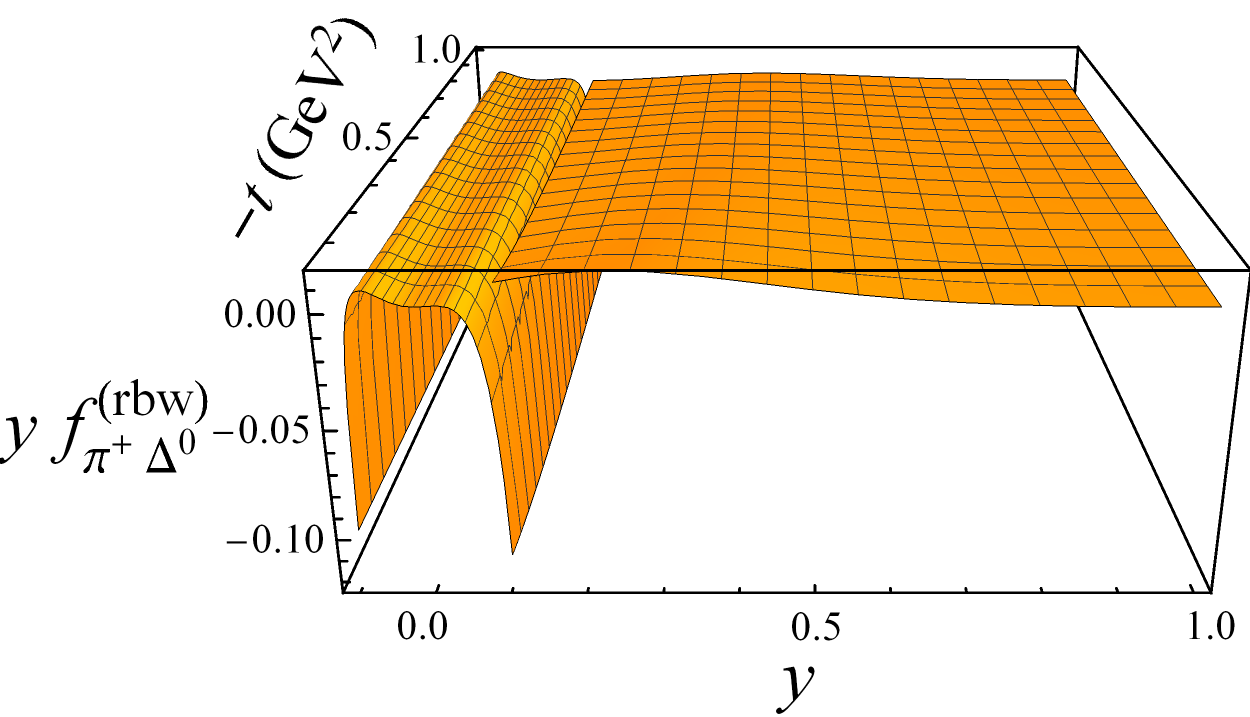}   
 \vspace{0pt}
\end{minipage}
\hfill
\begin{minipage}[t]{.45\linewidth}   
\hspace*{-0.85cm} \includegraphics[width=1.05\textwidth, height=5.5cm]{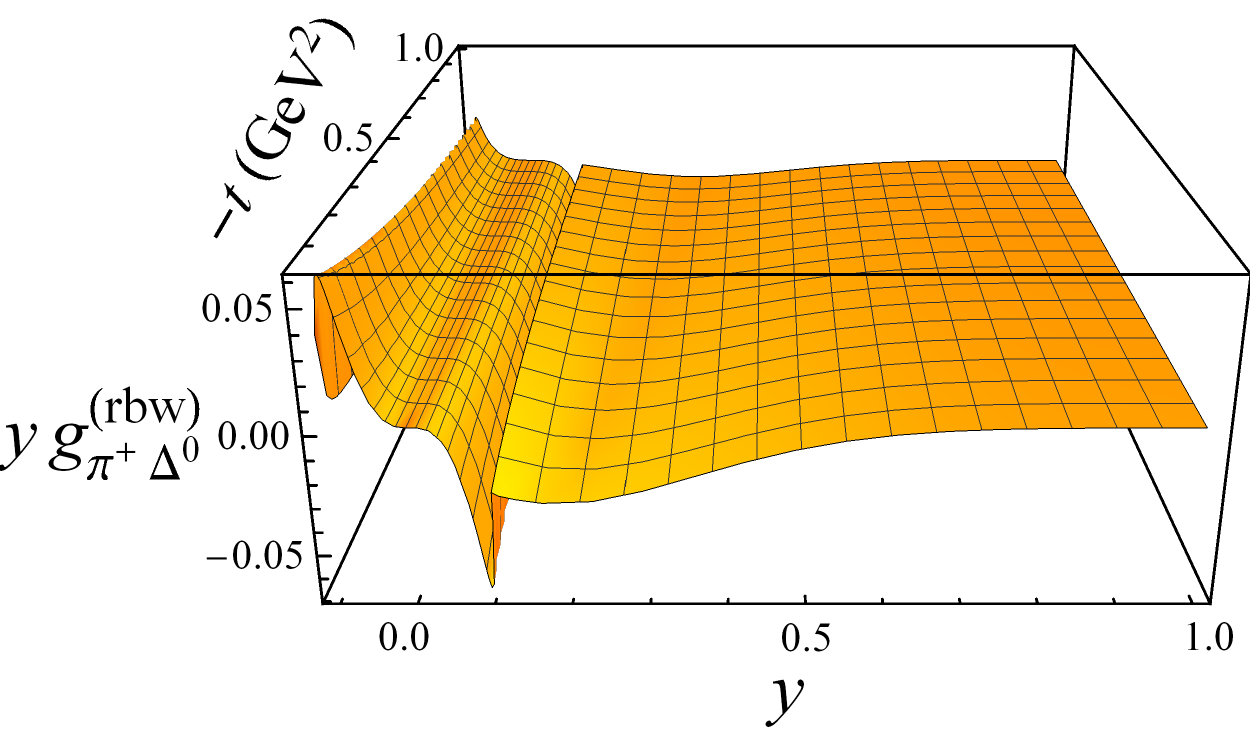}  
   \vspace{0pt}
\end{minipage} 
\begin{minipage}[t]{.45\linewidth}
\hspace*{-0.5cm}\includegraphics[width=1.1\textwidth, height=5.5cm]{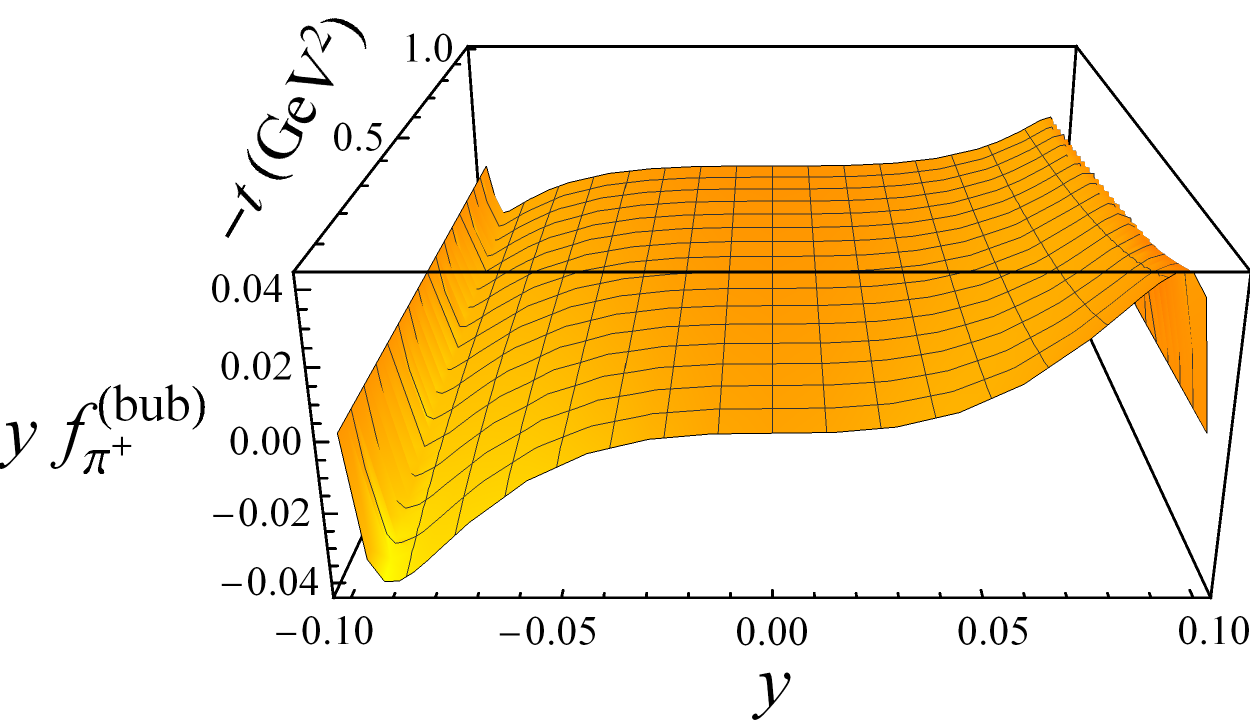}   
 \vspace{0pt}
\end{minipage}
\hfill
\begin{minipage}[t]{.45\linewidth}   
\hspace*{-0.85cm} \includegraphics[width=1.1\textwidth, height=5.5cm]{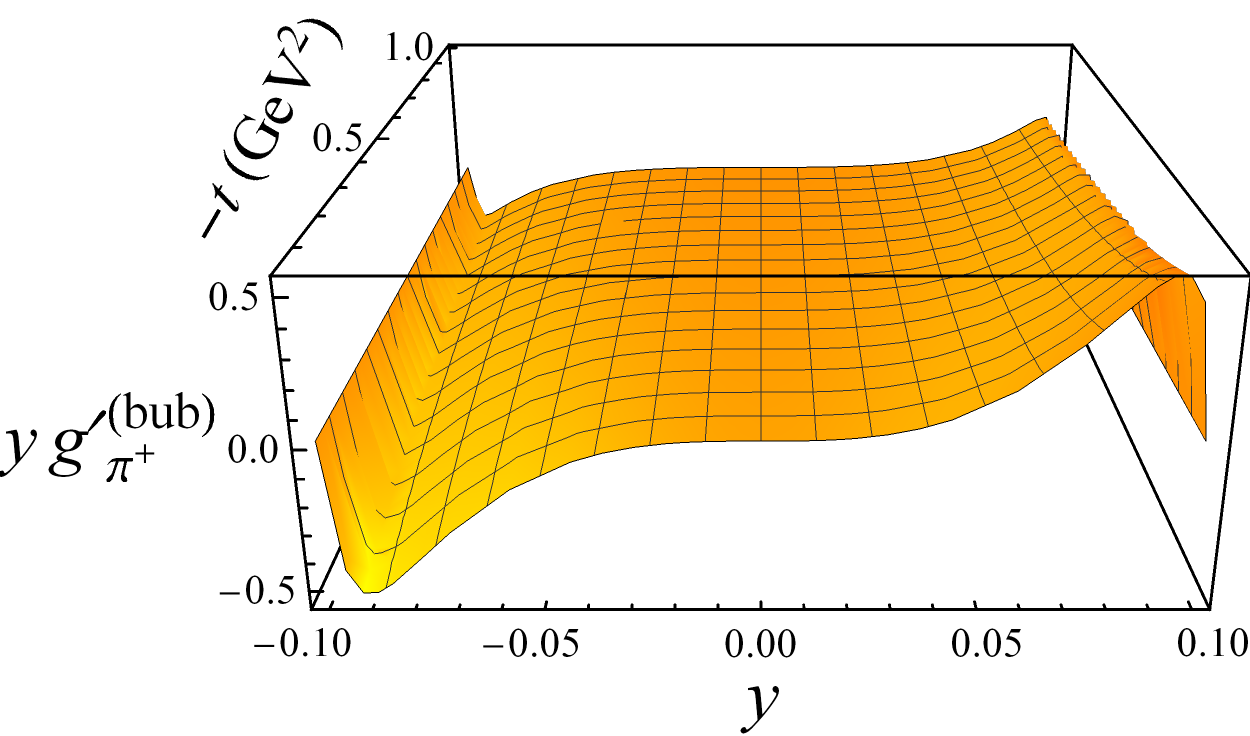}  
   \vspace{-12pt}
\end{minipage}
\caption{Three-dimensional splitting functions $y\, f_i(y,\xi,t)$ as a function of $y$ and $t$, with fixed $\xi=0.1$, for the octet baryon rainbow (upper panels), decuplet baryon rainbow (middle panels), bubble and magnetic bubble (lower panels) diagrams.}
\label{fig:sf_3d}
\end{figure}

The 3-dimensional splitting functions $y\, f_i(y,\xi,t)$ for the rainbow and bubble diagrams are shown in Fig.~\ref{fig:sf_3d}, where for the purpose of comparison we focus on the case of a charged $\pi^+$ loop. 
For illustration, we fix $\xi = 0.1$, and display the splitting functions for $y$ between $-\xi$ and 1, and for $-t < 1$~GeV$^2$.
As expected, the absolute values of these splitting functions decrease with increasing $-t$.
For the octet baryon rainbow diagram, the splitting function $y\, f_{\pi^+n}^{\rm (rbw)}$ has a similar peak value within the regions $y>\xi$ and $|y|<\xi$.
However, the absolute value of the Pauli splitting function $g_{\pi^+n}^{\rm (rbw)}$ in the $|y|<\xi$ region is smaller than that in the $y>\xi$ region. 
Both $f_{\pi^+n}^{\rm (rbw)}$ and $g_{\pi^+n}^{\rm (rbw)}$ are continuous at the boundary $y=\xi$. 
The shapes of these are similar in the $y>\xi$ region; however, the Dirac splitting function's shape is sharper when $y$ ranges from $-\xi$ to $\xi$.

For the decuplet baryon rainbow diagram, the shape of the splitting function $y\, g_{\pi^+ \Delta^0}^{\rm (rbw)}$ is similar as that of $y\, f_{\pi^+n}^{\rm (rbw)}$, although their signs in the $y>\xi$ region are opposite.
The Dirac-like splitting function for the decuplet rainbow $f_{\pi^+ \Delta^0}^{\rm (rbw)}$ decreases rapidly near the endpoints $y=\xi$ and $y=-\xi$, and is not continuous at the boundary $y=\xi$.
Such a discontinuity arises from terms with higher powers of the meson momentum in the loop integral for the splitting function, for example, the $(k \cdot p)^2 k\cdot p'$ terms in Eq.~(\ref{eq:decf_int}).
For the bubble diagrams in Figs.~\ref{fig:diagrams}(b) and \ref{fig:diagrams}(c), the relevant splitting functions are nonzero only when $y$ satisfies $|y|<\xi$, and $y\, f_{\pi^+}^{\rm (bub)}$ and $y\, g_{\pi^+}^{\prime{\rm (bub)}}$ are antisymmetric about the $y=0$ axis.
The absolute value of the Pauli-like $g_{\pi^+}^{\prime \rm (bub)}$ splitting function is approximately an order of magnitude larger than that of the Dirac-like function $f_{\pi^+}^{\rm (bub)}(y)$.

\begin{figure}[tbp]
\begin{minipage}[b]{.45\linewidth}
\hspace*{-0.3cm}\includegraphics[width=1.1\textwidth, height=5.5cm]{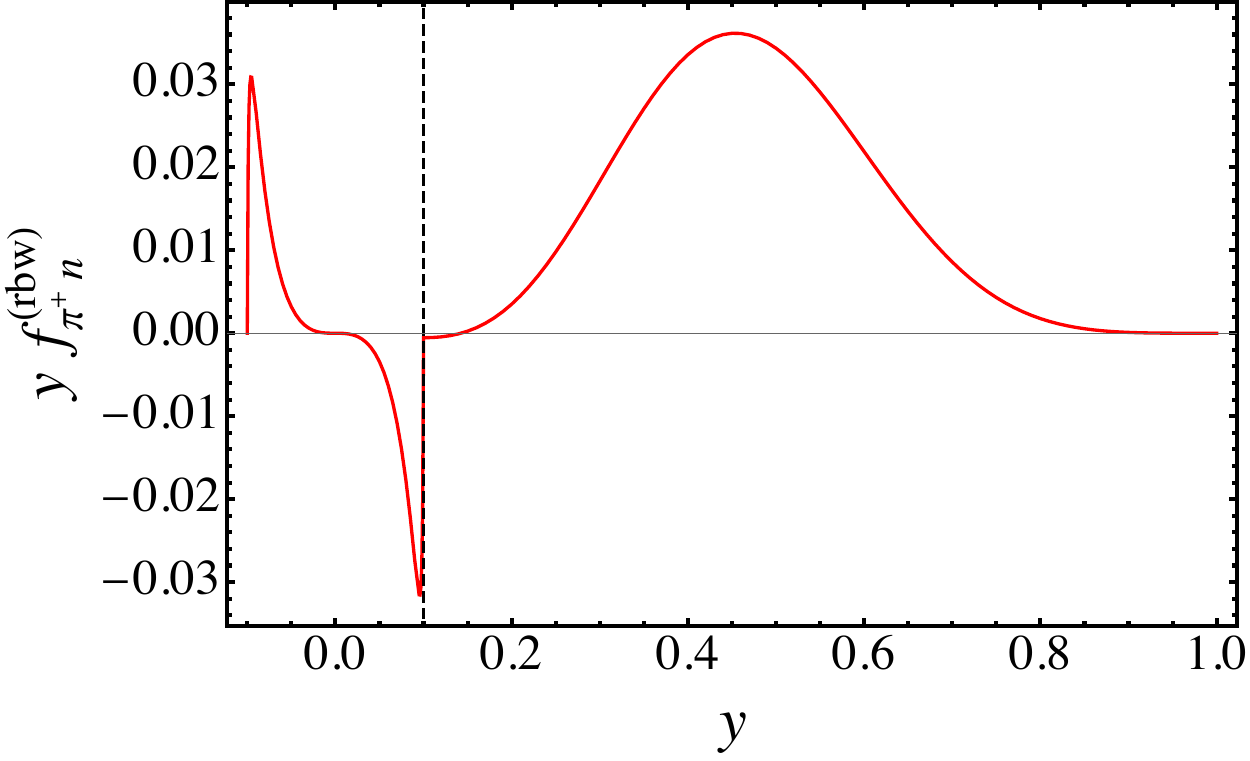}   
 \vspace{0pt}
\end{minipage}
\hfill
\begin{minipage}[b]{.45\linewidth}   
\hspace*{-0.95cm} \includegraphics[width=1.1\textwidth, height=5.5cm]{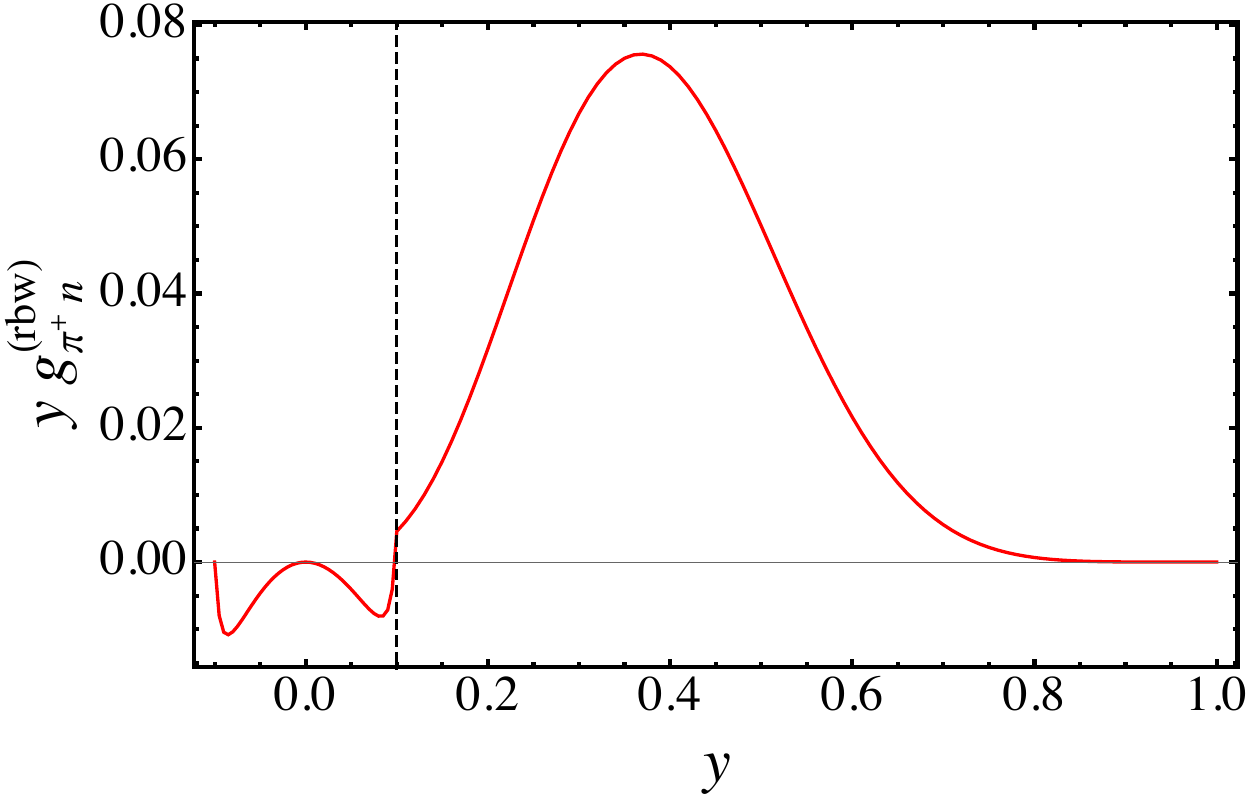} 
   \vspace{0pt}
\end{minipage}  
\\[-0.4cm]
\begin{minipage}[t]{.45\linewidth}
\hspace*{-0.5cm}\includegraphics[width=1.12\textwidth, height=5.5cm]{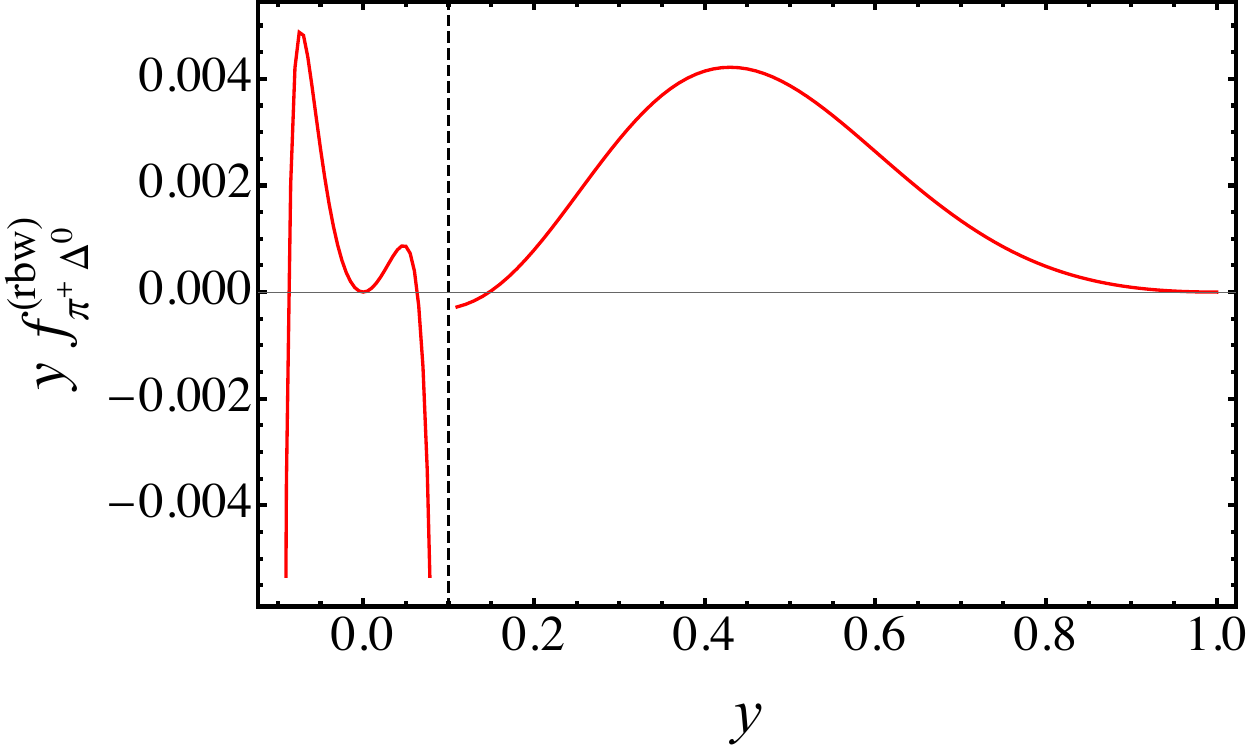}   
 \vspace{0pt}
\end{minipage}
\hfill
\begin{minipage}[t]{.45\linewidth}   
\hspace*{-0.85cm} \includegraphics[width=1.1\textwidth, height=5.5cm]{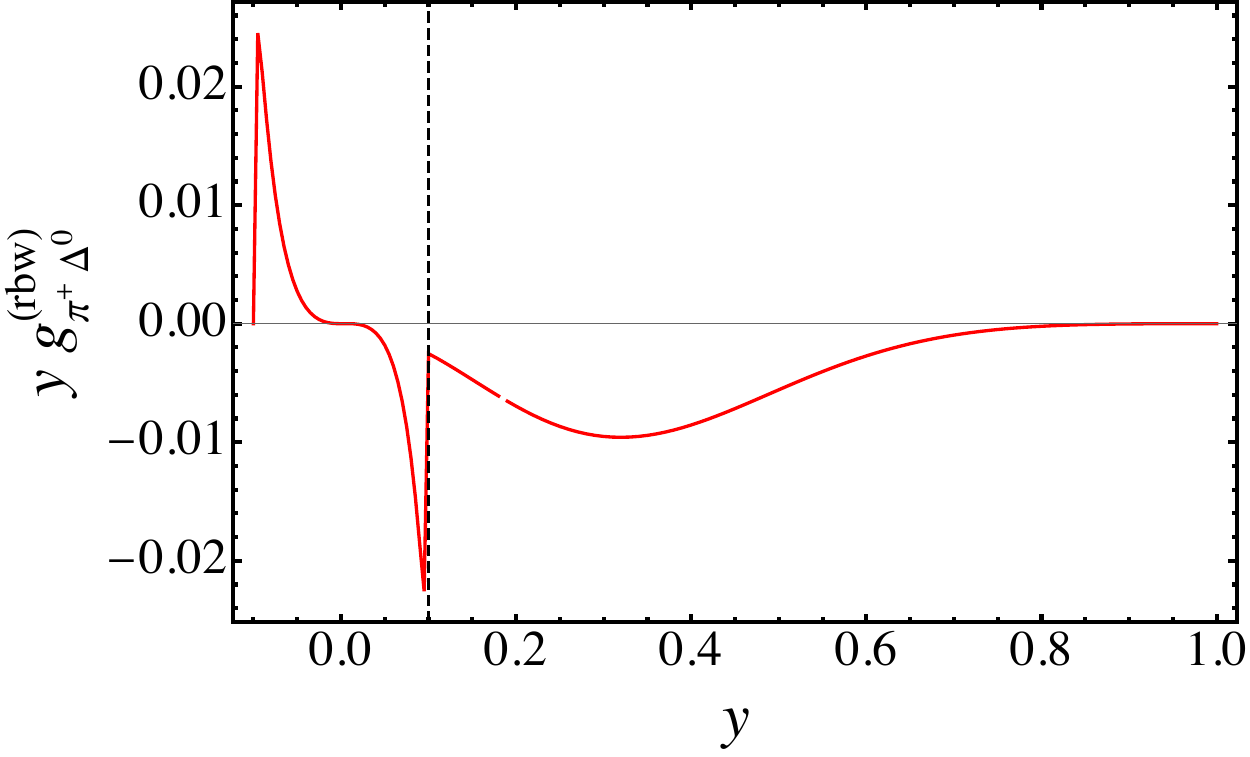}  
   \vspace{0pt}
\end{minipage} 
\begin{minipage}[t]{.45\linewidth}
\hspace*{-0.5cm}\includegraphics[width=1.12\textwidth, height=5.5cm]{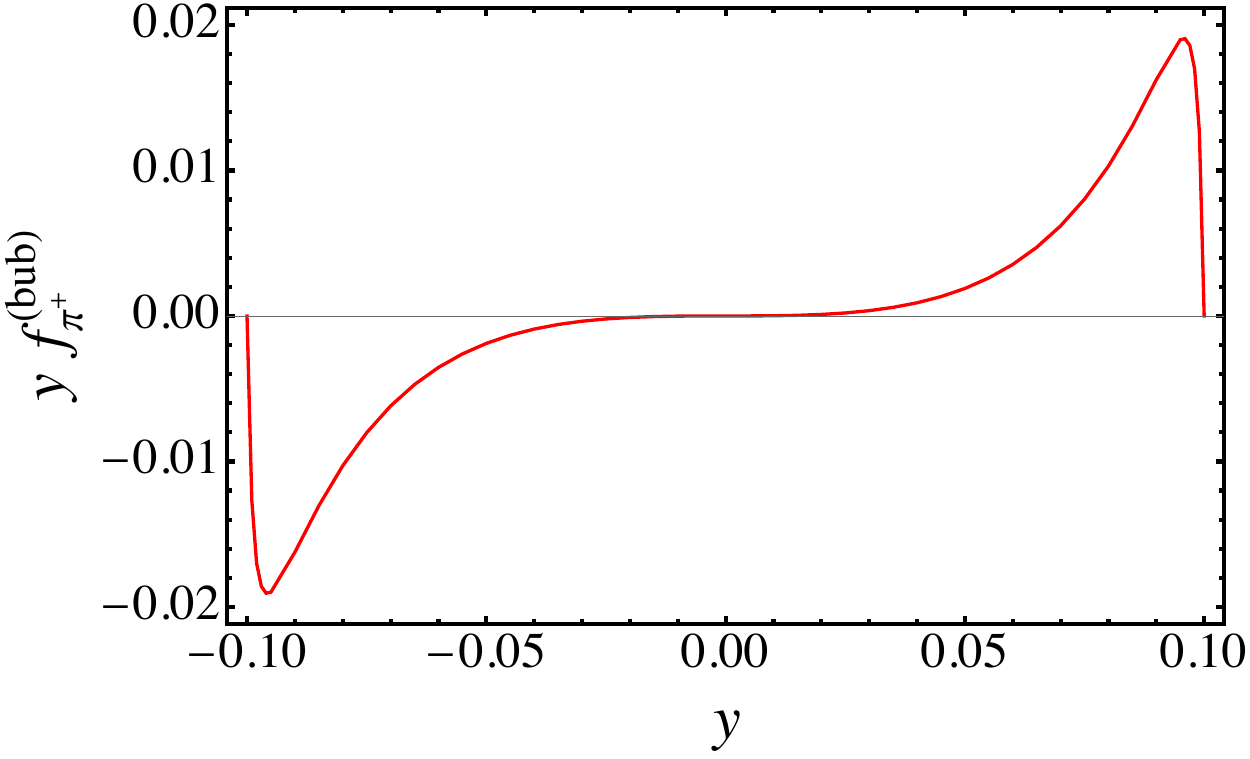}   
 \vspace{0pt}
\end{minipage}
\hfill
\begin{minipage}[t]{.45\linewidth}   
\hspace*{-0.85cm} \includegraphics[width=1.1\textwidth, height=5.5cm]{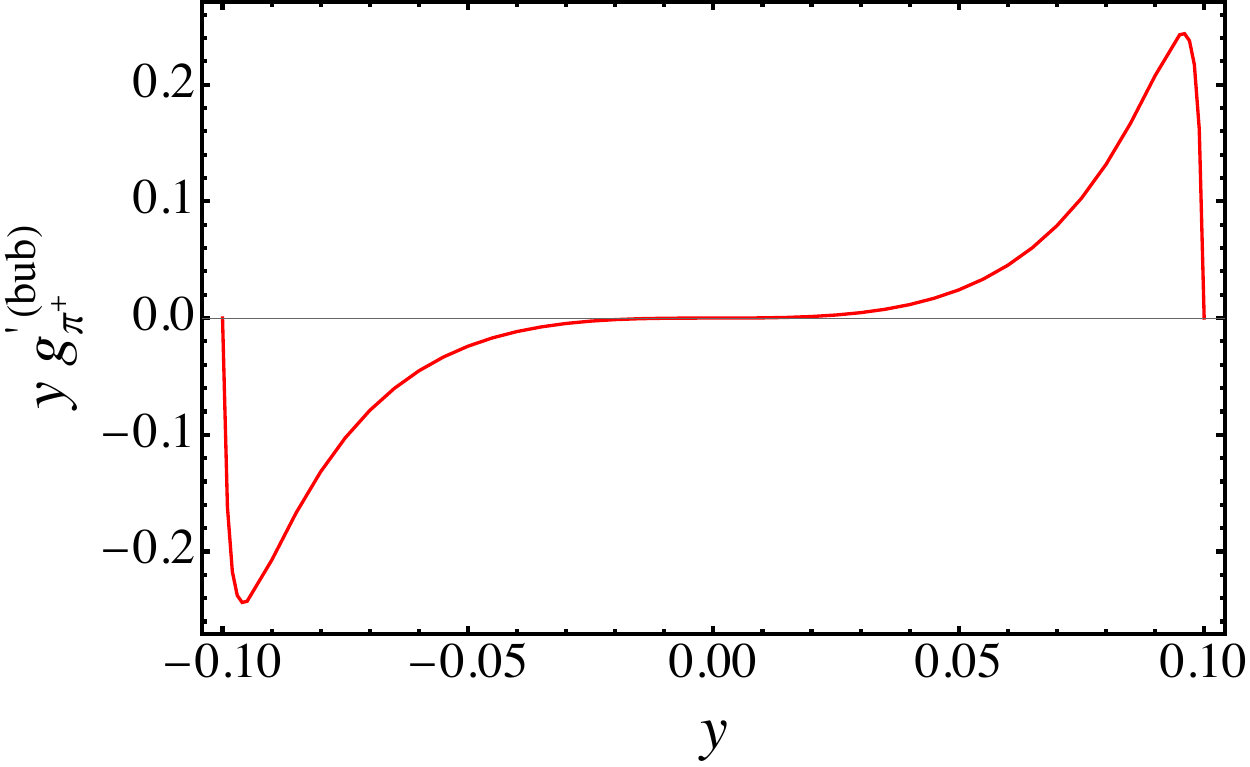}  
   \vspace{-12pt}
\end{minipage}
\caption{Two-dimensional projections of the Dirac-like $y f_i(y,\xi,t)$ and Pauli-like $y g_i(y,\xi,t)$ splitting functions for the octet baryon rainbow (top panels), decuplet baryon rainbow (middle panels), and bubble and magnetic bubble diagrams (lower panels), for $\xi=0.1$ and $-t=1$~GeV$^2$. The dashed vertical lines in the rainbow diagram plots correspond to the point $y = \xi$.}
\label{fig:sf_2d}
\end{figure}

To more clearly illustrate the shapes of the splitting functions, we show the two-dimensional projections of the functions in Fig.~\ref{fig:sf_2d} for $-t=1$~GeV$^2$ and $\xi=0.1$.
The splitting function $y\, f_{\pi^+n}^{\rm (rbw)}$ is approximately antisymmetric for $-\xi<y<\xi$.
Two sharp humps are seen when $y$ is close to the $\pm \xi$ points.
For the splitting function $g_{\pi^+n}^{\rm (rbw)}$, two humps are also seen when $-\xi<y<\xi$, although the absolute values of the maxima of the humps are smaller than those in the $\xi<y<1$ region.

For the decuplet baryon rainbow diagram, the two-dimensional splitting function $g_{\pi^+ \Delta^0}^{\rm (rbw)}$ has a similar shape to that of $f_{\pi^+n}^{\rm (rbw)}$. 
The shape of $f_{\pi^+ \Delta^0}^{\rm (rbw)}$ is different from that of the other functions when $-\xi<y<\xi$, with values near the $y=-\xi$ and $y=\xi$ boundary decreasing rapidly, and the shape is neither symmetric nor antisymmetric about the $y=0$ axis.
Such a behavior is in fact due to the effect from specific terms, which behave as $\delta(y)/y$ when $\xi$ tends to zero. 
The splitting functions for the bubble diagrams, $y\, f_{\pi^+}^{\rm (bub)}$ and $y\, g_{\pi^+}^{\prime \rm (bub)}$, are antisymmetric about the $y=0$ axis, so that by dividing by $y$ the resulting functions $f_{\pi^+}^{\rm (bub)}$ and $g_{\pi^+}^{\prime \rm (bub)}$ will be symmetric.

\begin{figure}[tbp]
\begin{minipage}[b]{.45\linewidth}
\hspace*{-0.3cm}\includegraphics[width=1.1\textwidth, height=5.5cm]{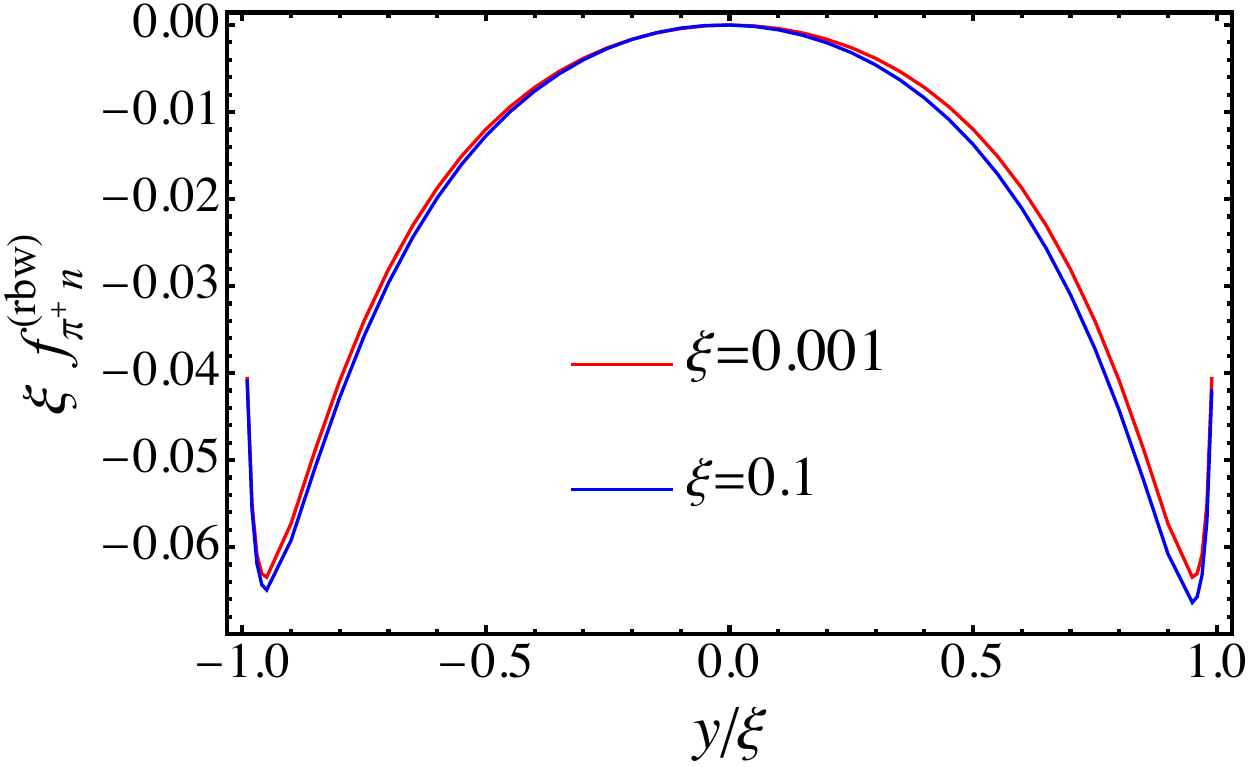}   
 \vspace{0pt}
\end{minipage}
\hfill
\begin{minipage}[b]{.45\linewidth}   
\hspace*{-0.95cm} \includegraphics[width=1.1\textwidth, height=5.5cm]{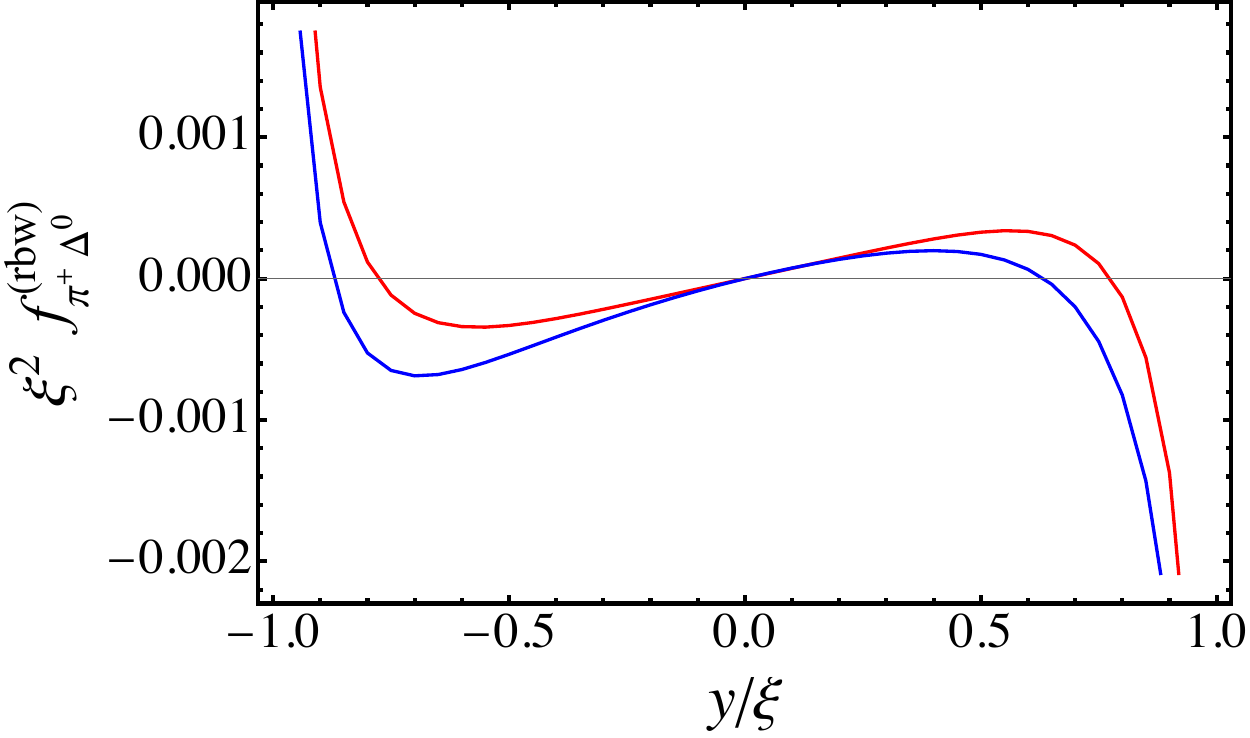} 
   \vspace{0pt}
\end{minipage}  
\\[-0.4cm]
\begin{minipage}[t]{.45\linewidth}
\hspace*{-0.5cm}\includegraphics[width=1.12\textwidth, height=5.5cm]{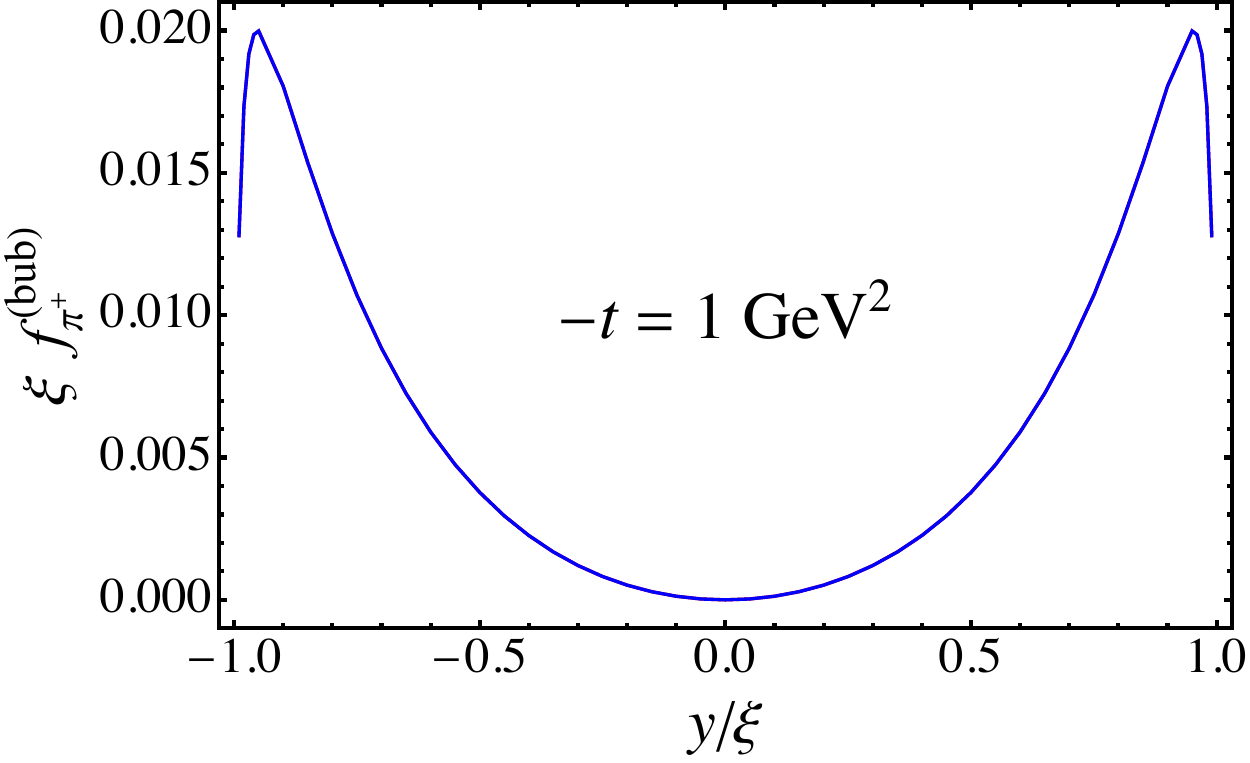} \vspace{0pt}
\end{minipage}
\hfill
\begin{minipage}[t]{.45\linewidth}   
\hspace*{-0.85cm} \includegraphics[width=1.1\textwidth, height=5.5cm]{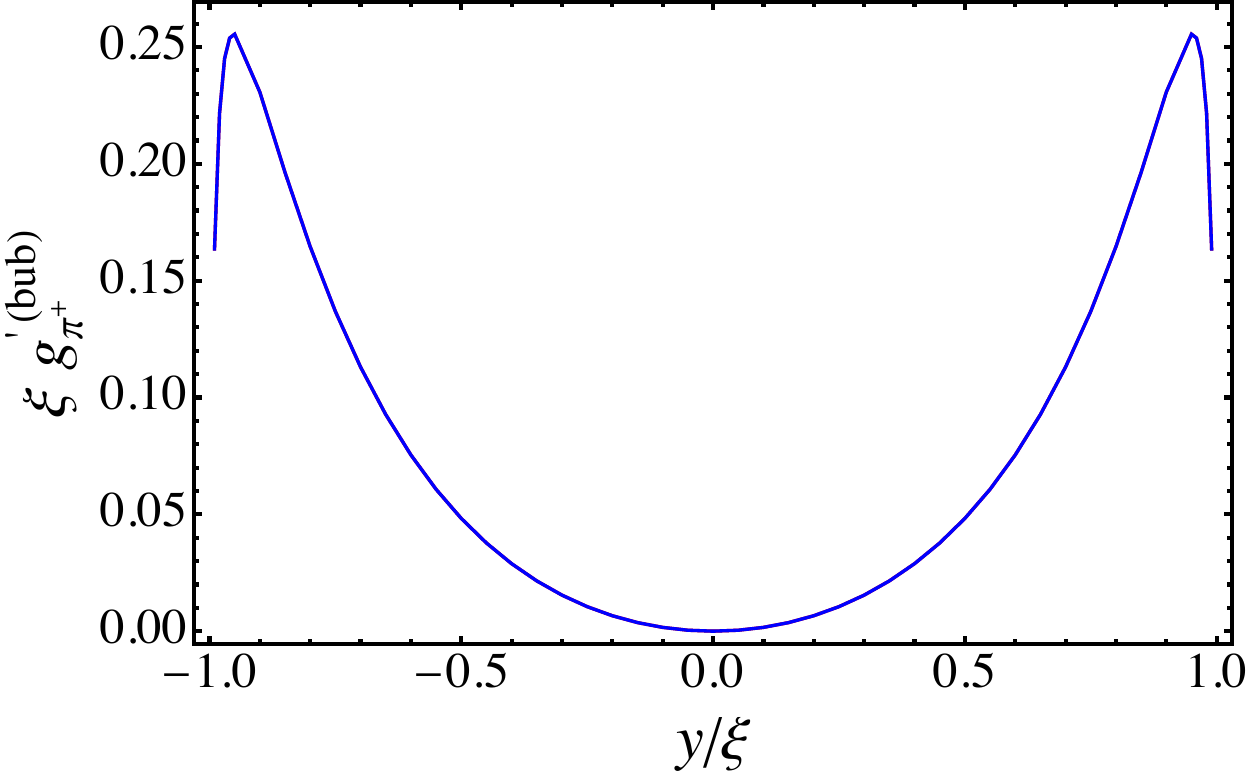}
   \vspace{-12pt}
\end{minipage} 
\caption{Scaled splitting functions $f_i(y,\xi,t)$ as a function of $y/\xi$ in the $-\xi<y<\xi$ region for $\xi=0.001$ (red lines) and $\xi=0.1$ (blue lines), for $-t = 1$~GeV$^2$.  All the functions are scaled by a factor of $\xi$, except for the decuplet function $f_{\pi^+ \Delta^0}^{\rm (rbw)}(y,\xi,t)$, which is scaled by a factor $\xi^2$.}
\label{fig:sf_erbl}
\end{figure}

In Fig.~\ref{fig:sf_erbl}, we show the results for the splitting functions $f_{\pi^+ n}^{\rm (rbw)}$, $f_{\pi^+ \Delta^0}^{\rm (rbw)}$, $f_{\pi^+}^{\rm (bub)}$, and $g_{\pi^+}^{\prime \rm (bub)}$ when $y$ is in the region $[-\xi,\xi]$, for $\xi$ values between 0.001 and 0.1.
To more conveniently compare the results with different $\xi$, we scale the horizontal axis by dividing $y$ by $\xi$, and the vertical axis by multiplying the splitting functions by $\xi$.
The scaled results with different $\xi$ values are very similar and converge as $\xi$ becomes small, with the results for $\xi<0.001$ almost indistinguishable from those at $\xi=0.001$.
This behavior is reminiscent of the behavior of a $\delta$-function, as has been discussed in previous chiral loop calculations~\cite{Salamu:2014pka, Salamu:2018cny}. 
In contrast, for the decuplet baryon splitting function one observes convergence at small $\xi$ after multiplying $f_{\pi^+\Delta^0}$ by $\xi^2$.
Since it is antisymmetric in $y$, this indicates that the splitting function will approach $\delta(y)/y$ rather than $\delta(y)$ in the zero-skewness limit.

\subsection{Pion loop contributions to quark GPDs}
\label{sec:res_GPDs}
\begin{figure}[tbp]
\begin{minipage}[b]{.45\linewidth}
\hspace*{-0.3cm}\includegraphics[width=1.1\textwidth, height=5.5cm]{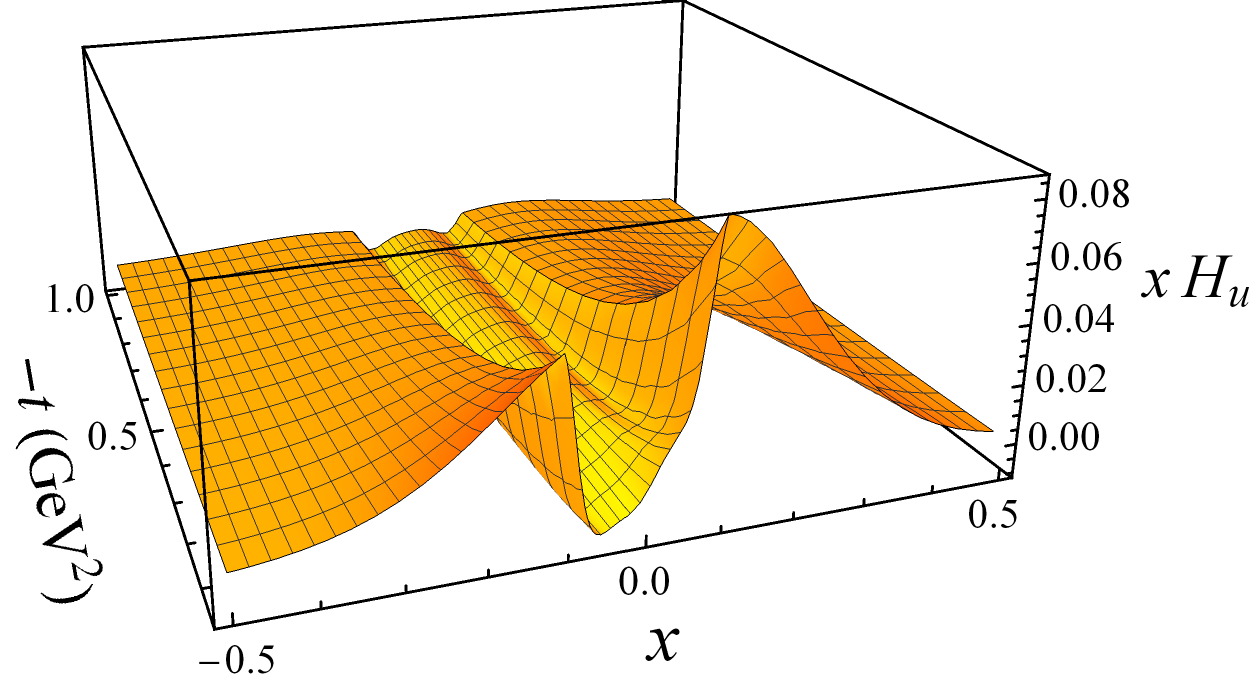}  \vspace{0pt}
\end{minipage}
\hfill
\begin{minipage}[b]{.45\linewidth}   
\hspace*{-0.95cm} \includegraphics[width=1.1\textwidth, height=5.5cm]{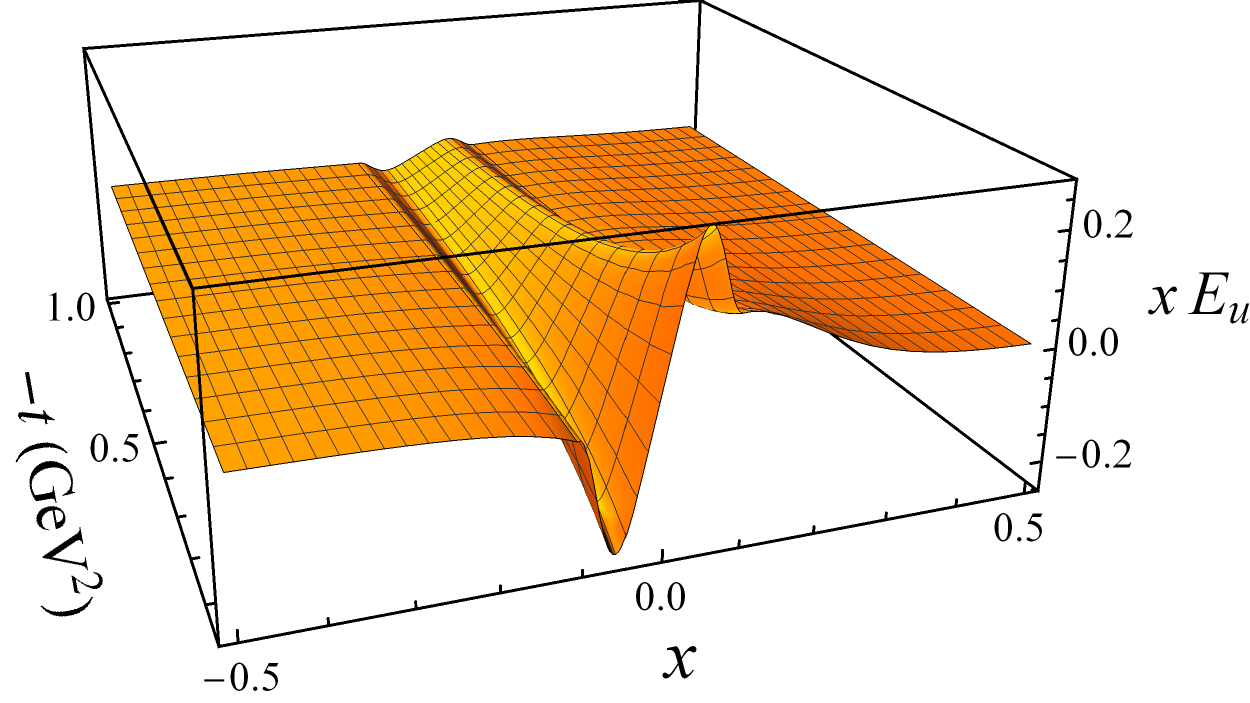} 
\vspace{0pt}
\end{minipage}  
\\[-0.4cm]
\caption{Total three-dimensional $u$-quark GPDs $x H_u$ and $x E_u$ from Eq.~(\ref{eq.Hu_alldiags}) as a function of $x$ and $t$, for fixed $\xi=0.1$ and energy scale $\mu=0.63$~GeV. The corresponding $d$-quark distributions can be obtained from the $u$-quark GPDs using the isospin symmetery relation, $\{H,E\}_d(x,\xi,t) = -\{H,E\}_u(-x,\xi,t)$, which holds for the contributions from virtual pion loops as in Fig.~\ref{fig:diagrams}.}
\label{fig:GPD_3d}
\end{figure}

Performing the convolutions of the splitting functions with the corresponding quark GPDs of the virtual pions, we show the contributions of the chiral loop to the $H$ and $E$ GPDs for $u$ quarks in Fig.~\ref{fig:GPD_3d} as a function of $x$ and $t$ at a fixed $\xi=0.1$.
For the contributions from the couplings to the virtual pion loops, as in Fig.~\ref{fig:diagrams}, the results for the $d$-quark GPDs can be obtained from the $u$-quark distributions using isospin symmetry, $\{H,E\}_d(x,\xi,t)=-\{H,E\}_u(-x,\xi,t)$.
The most striking structure is seen in the region of low $x$, $|x| \lesssim 0.2$.
For the case of the electric $u$-quark GPD, $x H_u$ is positive in the DGLAP region, and has two valleys in the ERBL region.
For the magnetic GPD, $x E_u$ is negative in the $x < 0$ region and positive when $x > 0$.
The distributions fall rapidly as $|x| \to 1$, and for increasing values of the momentum transfer squared, $-t$.

\begin{figure}[tbp]
\begin{minipage}[b]{.45\linewidth}
\hspace*{-0.3cm}\includegraphics[width=1.1\textwidth, height=5.5cm]{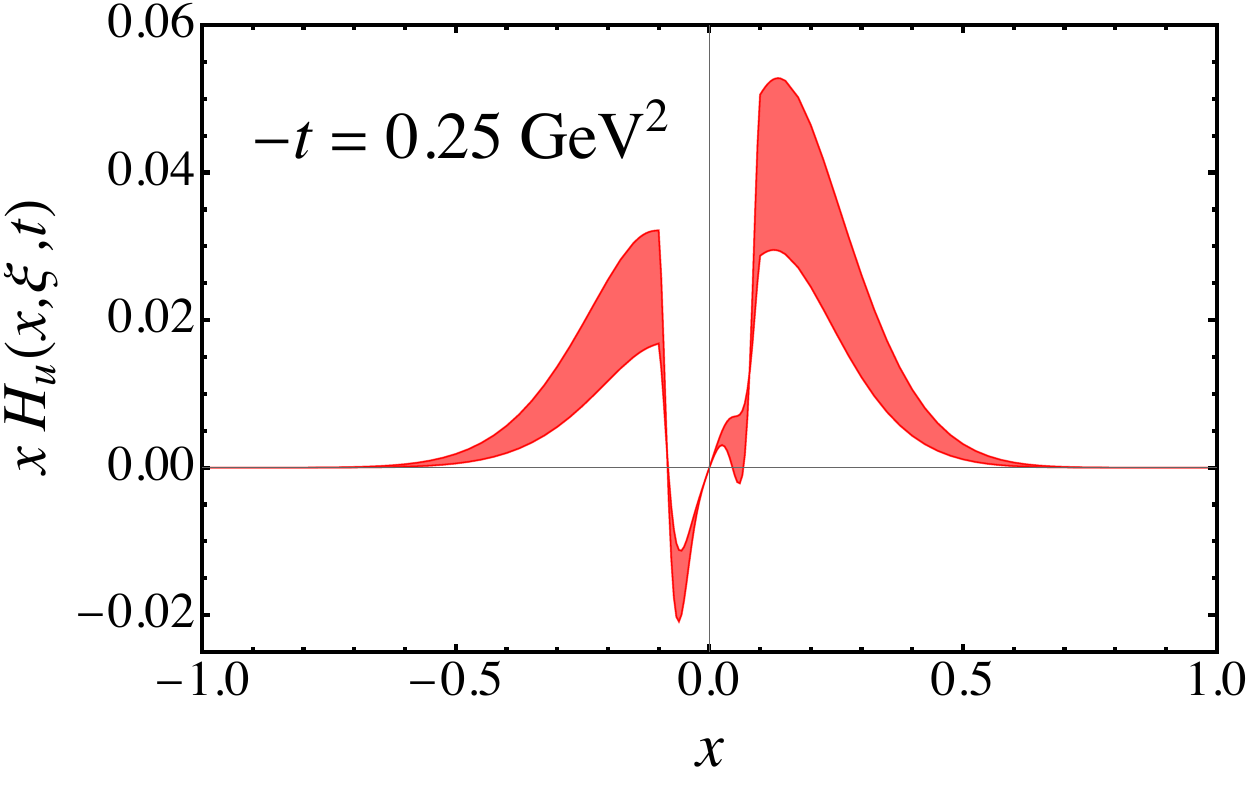}   
 \vspace{0pt}
\end{minipage}
\hfill
\begin{minipage}[b]{.45\linewidth}   
\hspace*{-0.95cm} \includegraphics[width=1.1\textwidth, height=5.5cm]{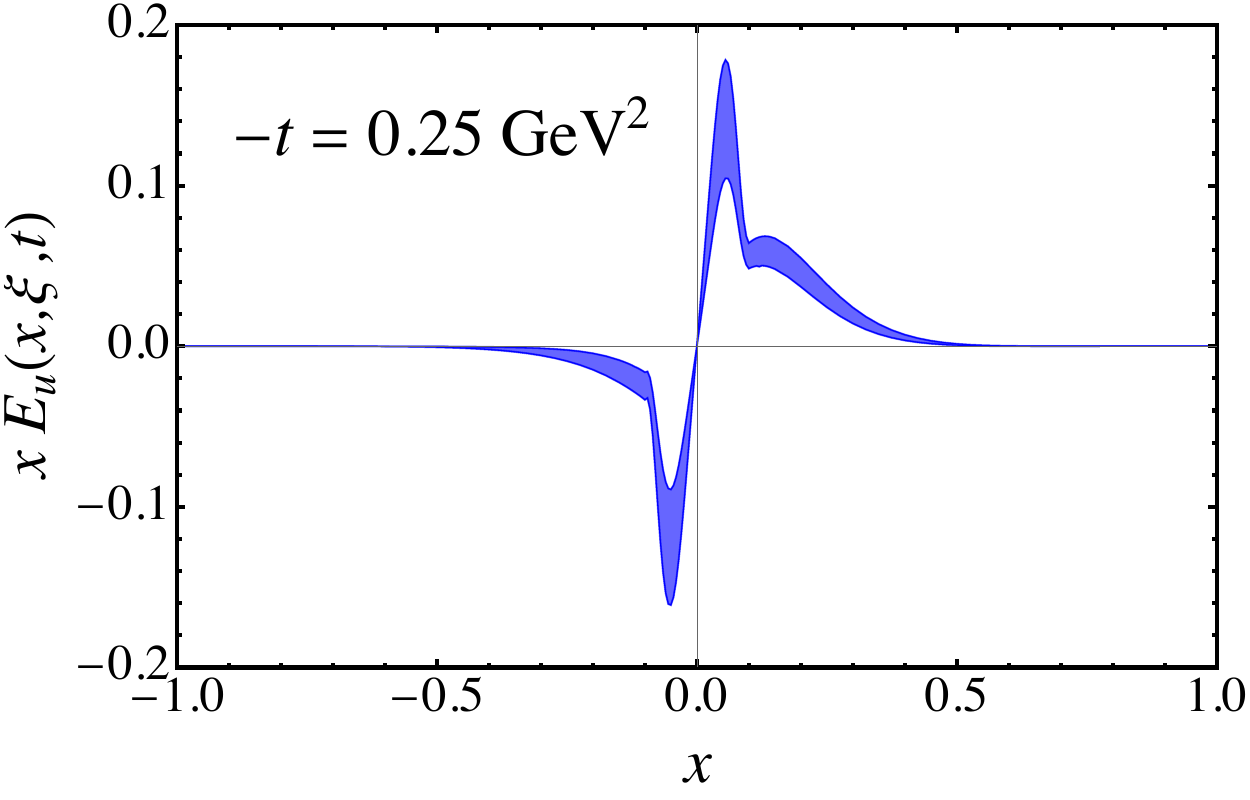} 
   \vspace{0pt}
\end{minipage}  
\\[-0.4cm]
\begin{minipage}[t]{.45\linewidth}
\hspace*{-0.5cm}\includegraphics[width=1.12\textwidth, height=5.5cm]{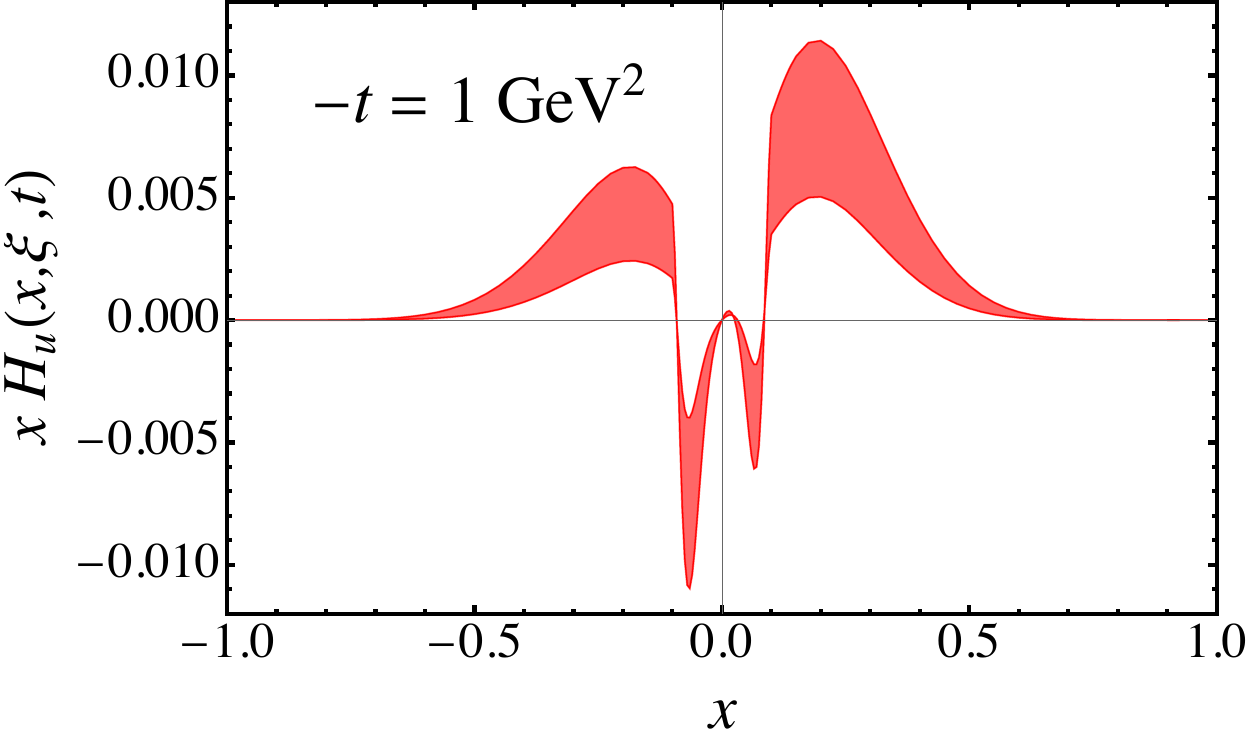}   
 \vspace{0pt}
\end{minipage}
\hfill
\begin{minipage}[t]{.45\linewidth}   
\hspace*{-0.85cm} \includegraphics[width=1.1\textwidth, height=5.5cm]{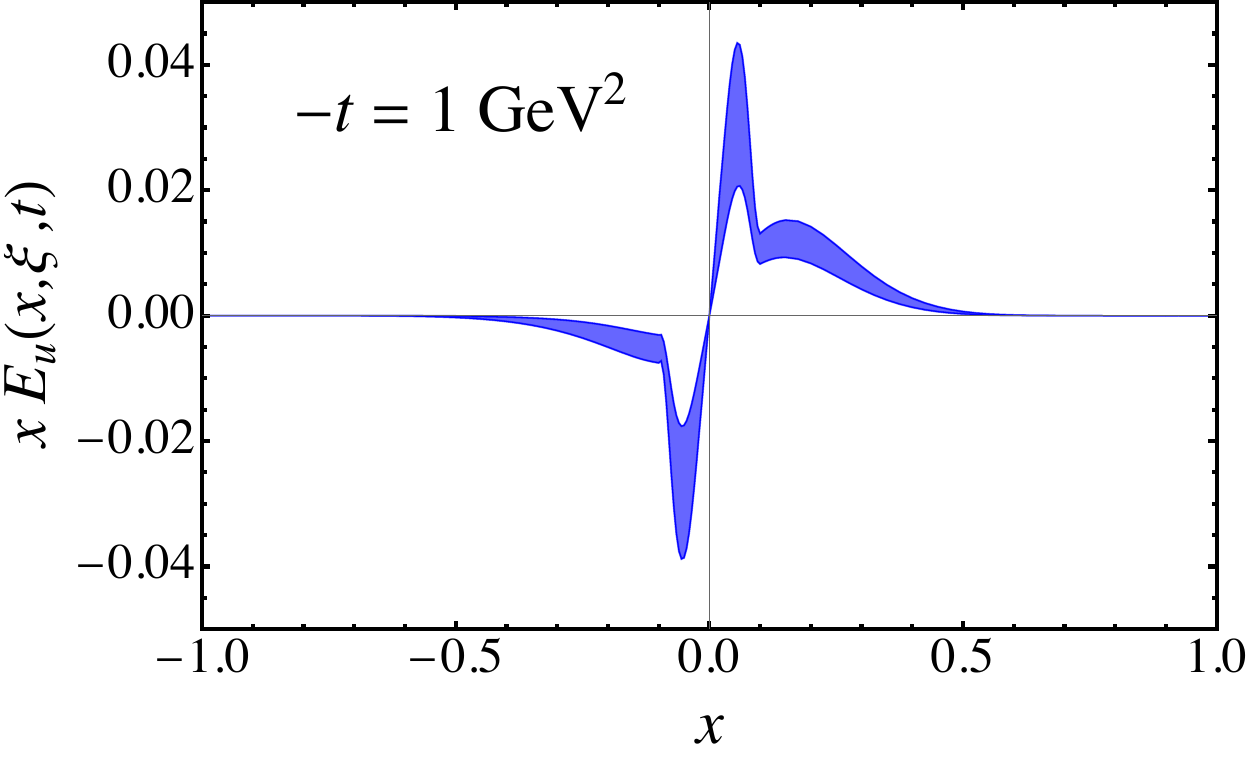}  
   \vspace{-12pt}
\end{minipage} 
\caption{Meson loop contributions to the GPDs $H_u$ and $E_u$ for $\xi=0.1$ and $-t=1$~GeV$^2$ at the energy scale $\mu=0.63$~GeV. The corresponding $d$-quark GPDs can be obtained from the relation $\{H,E\}_{d}(x,\xi,t)=-\{H,E\}_{u}(-x,\xi,t)$.}
\label{fig:GPD_2d_1GeV}
\end{figure}

To more clearly illustrate the dependence of the GPDs on the parton momentum fraction~$x$, in Fig.~\ref{fig:GPD_2d_1GeV} we plot the two-dimensional projections of the GPDs $x H_u$ and $x E_u$ for $-t=0.25$~GeV$^2$ and $-t=1$~GeV$^2$. 
Although the splitting function $f_{\pi^+\Delta^0}^{\rm (rbw)}$ is discontinuous at $y=\xi$, as evident from Fig.~\ref{fig:sf_2d}, the quark GPDs remain continuous at $x=\pm\xi$. 
The reason is that the convolution formulas in Eqs.~(\ref{eq:conv_a}), (\ref{eq:conv_b}) and (\ref{eq:conv_d}) are only related to the splitting function in the $y>\xi$ region, and the results obtained using these should be continuous at $\xi$ if the input pion GPD is continuous at $x=\xi$, which guarantees the the amplitudes for DVCS and HEMP
are finite~\cite{Diehl:2003ny}. 
However, there are no constraints on the derivatives of GPDs at $x=\xi$, and the discontinuity of the derivative can arise from the different integration regions in the $\alpha$--$\beta$ plane for the DGLAP and ERBL regions when using the double distribution 
parametrization in Eq.~(\ref{eq:inputGPD})~\cite{Musatov:1999xp, Diehl:2003ny}. 
As the contribution of $D$-term in Eq.~(\ref{eq:inputGPD}) only exists in the ERBL region, it also leads to a discontinuity of the first derivative of the GPD at $x=\xi$.
On the other hand, the contribution to the quark GPD from Eq.~(\ref{eq:conv_c}) vanishes $x=\xi$ because of the endpoint property of the distribution amplitude, $\Phi_{q/\pi}(1,\kappa,s)=0$.

Comparing the results with different $-t$, the absolute values of the GPDs at $-t = 0.25$~GeV$^2$ are around 4--5 times larger than those at $-t = 1$~GeV$^2$. 
For the $u$-quark distribution, one finds that the quark GPD $H_u$ in the DGLAP region for $x>0$ has a larger magnitude than $H_u$ at $x<0$ (which by crossing symmetry is equivalent to the $\bar u$ distribution at $x>0$).
According to the above isospin relation for the contributions from the pion coupling diagrams, $xH_u(x>0)$ is identical to $xH_d(x<0)$, the latter which is equivalent to the $x\bar d$ distribution at $x>0$.
Therefore the contributions from the loop diagrams in Fig.~\ref{fig:diagrams} naturally give an enhancement of the $\bar d$ distribution compared with the $\bar u$, reminiscent of the empirical result for the $\bar d$ and $\bar u$ PDF asymmetry in the collinear region~\cite{Salamu:2014pka}.

Moreover, the result in the $0<x<\xi$ region is positive for $-t = 0.25$~GeV$^2$, but negative for $-t = 1$~GeV$^2$, which can be understood from the fact that the pion GPD or GDA includes the $D_{q/\pi}(t)$ form factor in the ERBL region that has a different $t$ dependence compared with the pion form factor $F_\pi(t)$.
The magnitude of the Pauli GPD $E_u$ is larger than that of the Dirac GPD $H_u$, and the quark and antiquark distributions in the DGLAP region have opposite signs.
This implies that the GPD for the $\bar u$ flavor has a different sign to that of the Dirac GPD for $\bar d$, and the absolute value of the Pauli GPD $E$ for $\bar d$ is larger than that for $\bar u$.

\subsection{Form factors}

From the calculated $H_q$ and $E_q$ GPDs one can compute the Dirac, Pauli, and gravitational form factors from the lowest two moments of the GPDs.
The Dirac and Pauli form factors can be obtained from Eq.~(\ref{eq.F1q}) and (\ref{eq.F2q}), respectively.
For the gravitational form factors, the pion loop contributions to $A^q$, $B^q$ and $C^q$ can be written as
\begin{subequations}
\label{eq.gravFFs}
\begin{eqnarray}
A^q(t) &=&\sum_i 
A^{2(0)}_{q/\pi}(t)\, {\cal A}^{2(0)}_{i}(t),
\\
B^q(t) &=&\sum_i 
A^{2(0)}_{q/\pi}(t)\, {\cal B}^{2(0)}_{i}(t),
\\
C^q(t) &=&\sum_i 
\Big[ A^{2(0)}_{q/\pi}(t)\, {\cal C}^{(2)}_i(t) 
    + C^{(2)}_{q/\pi}(t)\,  {\cal C}'_i(t) 
\Big],
\end{eqnarray}
\end{subequations}
where $A^{2(0)}_{q/\pi}$ and $ C^{(2)}_{q/\pi}$ are the quark gravitational form factors in the pion, obtained from the pion GPD in Eq.~(\ref{eq:inputGPD}) through
\begin{equation}
\int_{-1}^1 \dd{x} x\, H_{q/\pi}(x,\xi,t)
= A^{2(0)}_{q/\pi}(t) + (2\xi)^2\, C^{(2)}_{q/\pi}(t).
\end{equation}
Note that the $C^{(2)}_{q/\pi}$ form factor is proportional to the pion gravitational form factor $D_{q/\pi}(t)$ in Eq.~(\ref{eq:Dpit}),
    $C^{(2)}_{q/\pi}(t) = \frac14 D_{q/\pi}(t)$.
The sum over the index $i$ in Eqs.~(\ref{eq.gravFFs}) includes the various hadronic configurations in Fig.~\ref{fig:diagrams}, and the corresponding hadronic form factors ${\cal A}^{2(0)}_i$, ${\cal B}^{2(0)}_i$ and ${\cal C}_i^{(2)}$ are obtained from the first moments of splitting functions, 
\begin{subequations}
\begin{eqnarray}
\int_{-\xi}^1 \dd{y}\, y\, f_i(y,\xi,t)
&=& {\cal A}^{2(0)}_i(t) + (2\xi)^2\, {\cal C}^{(2)}_i(t),
\\
\int_{-\xi}^1 \dd{y}\, y\, g_i(y,\xi,t)
&=& {\cal B}^{2(0)}_i(t) - (2\xi)^2\, {\cal C}^{(2)}_i(t),
\end{eqnarray}
\end{subequations}
with the form factor ${\cal C}'_i(t)$ for the rainbow diagrams in Figs.~\ref{fig:diagrams}(a) and \ref{fig:diagrams}(d) obtained from 
\begin{eqnarray}
\label{eq:Cprime}
{\cal C}'_i(t) 
&=& \int_{-\xi}^1 \frac{\dd{y}}{y}\,\, f_i(y,\xi,t)
 = -\int_{-\xi}^1 \frac{\dd{y}}{y}\,\, g_i(y,\xi,t).
\end{eqnarray}
The ${\cal C}'_i(t)$ form factor for Figs.~\ref{fig:diagrams}(b) and \ref{fig:diagrams}(c) can also be obtained from Eq.~(\ref{eq:Cprime}) by changing the integral region to $y\in[-\xi,\xi]$. 
However, since for these diagrams the functions $\frac{1}{y}f_{\pi^+}^{\rm (bub)}$ and $\frac{1}{y}g_{\pi^+}'^{\rm (bub)}$ are antisymmetric with respect to $y=0$, the integrals in Eq.~(\ref{eq:Cprime}), and hence ${\cal C}'_i(t)$, vanish.
(Note that the functions $y f_{\pi^+}^{\rm (bub)}$ and $y g_{\pi^+}'^{\rm (bub)}$ shown in Fig.~\ref{fig:sf_2d} are also antisymmetric with respect to $y$.)

\begin{figure}[tbp] 
\begin{minipage}[b]{.45\linewidth}
\hspace*{-2.3cm}\includegraphics[width=1.5\textwidth, height=7.5cm]{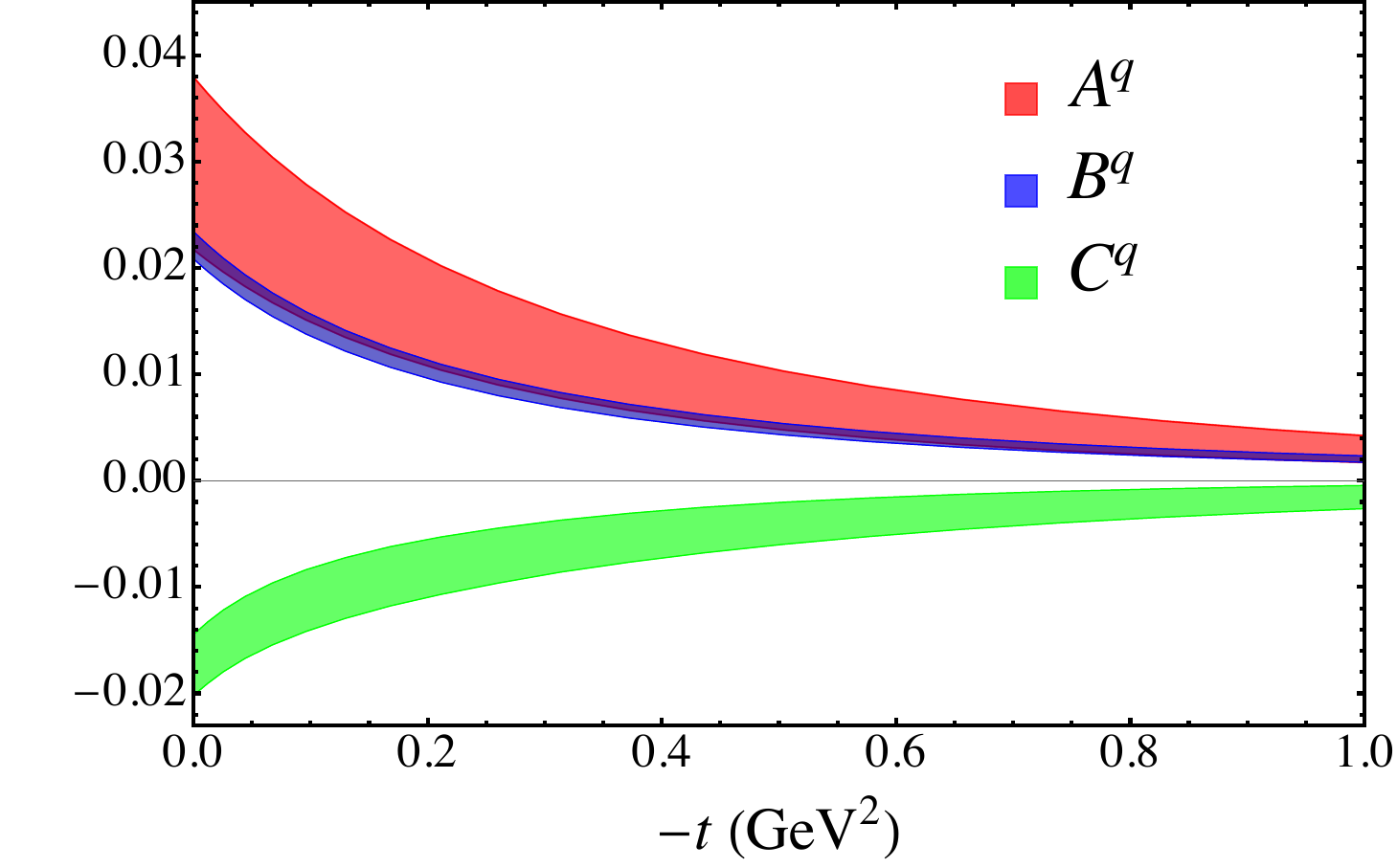}   
\end{minipage}
\caption{Gravitational form factors $A^q$ (red band), $B^q$ (blue band), and $C^q$ (green band) for quark flavor $q=u$ or $d$ as a function of $t$ arising from the pseudoscalar meson loop diagrams in Fig.~\ref{fig:diagrams}. The results for the $u$ and $d$ flavors for these diagrams are identical due to isospin symmetry.}
\label{fig:GFFs_ud}
\end{figure}

The numerical results for the gravitational $A^q$, $B^q$ and $C^q$ form factors are shown in Fig.~\ref{fig:GFFs_ud} as a function of $t$.
Because of isospin symmetry the results for the $u$ 
flavor are identical to those for the $d$ 
flavor.
Despite the absolute value of the GPD $E_q$ being larger than that of the GPD $H_q$, the first moment of $E_q$ largely cancels between the positive-$x$ and negative-$x$ regions, leading to similar values for the $A^q$ and $B^q$ form factors.
In contrast, the $C^q$ form factor is negative and has a smaller magnitude compared with the other two form factors.

As pointed out in the preceeding discussions, in this analysis we only consider contributions to GPDs arising from the meson loop diagrams in Fig.~\ref{fig:diagrams}. 
Generally, comparison with experimental measurements or other phenomenological calculations of the total quark gravitational form factors will require inclusion of contributions from coupling to intermediate state baryons in the loops or Kroll-Ruderman type contributions, which will be considered in future analyses.
On the other hand, in a valence quark approximation the antiquark distributions arise predominantly from couplings to the meson loop, which allows their contributions to form factors and sea quark flavor asymmetries to be studied directly in terms of the diagrams in Fig.~\ref{fig:diagrams} alone.

\begin{figure}[tbp] 
\begin{minipage}[b]{.45\linewidth}
\hspace*{-0.3cm}\includegraphics[width=1.1\textwidth, height=5.5cm]{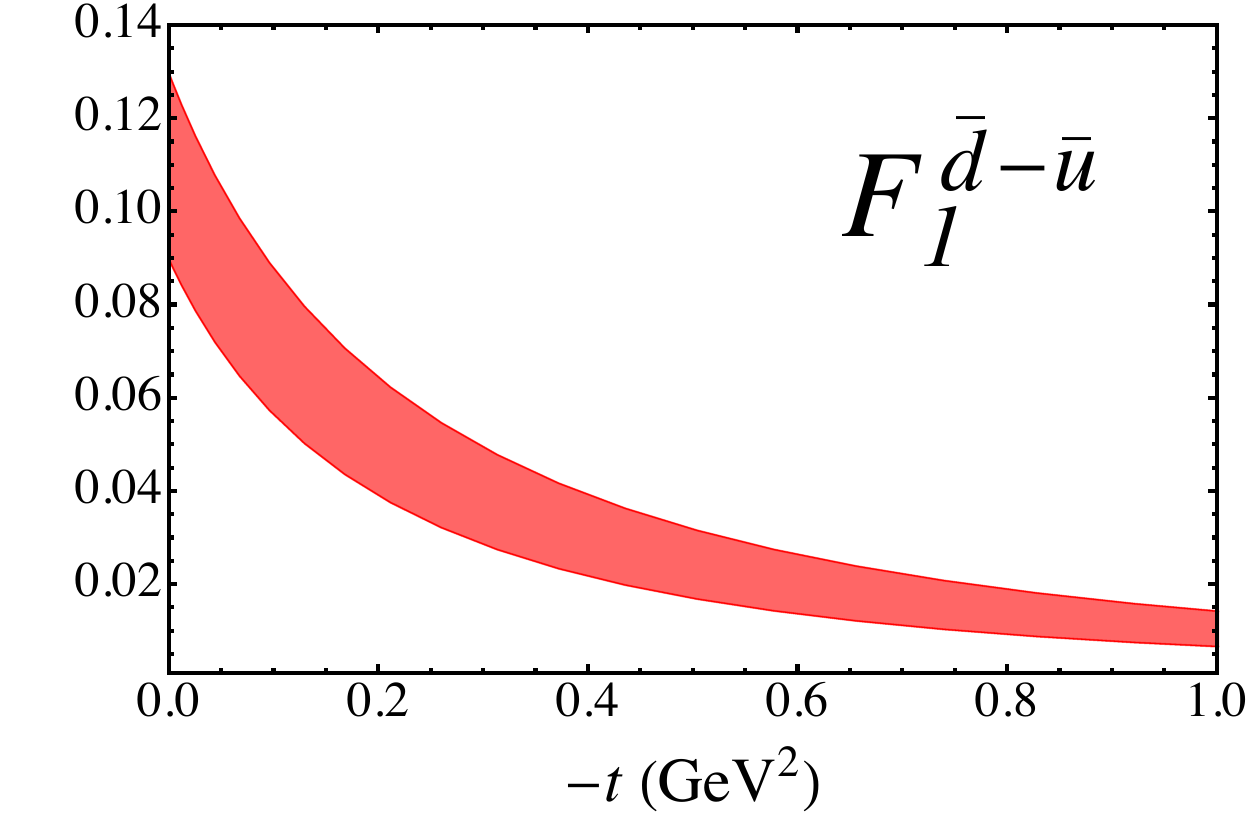}   
 \vspace{0pt}
\end{minipage}
\hfill
\begin{minipage}[b]{.45\linewidth}   
\hspace*{-0.95cm} \includegraphics[width=1.1\textwidth, height=5.5cm]{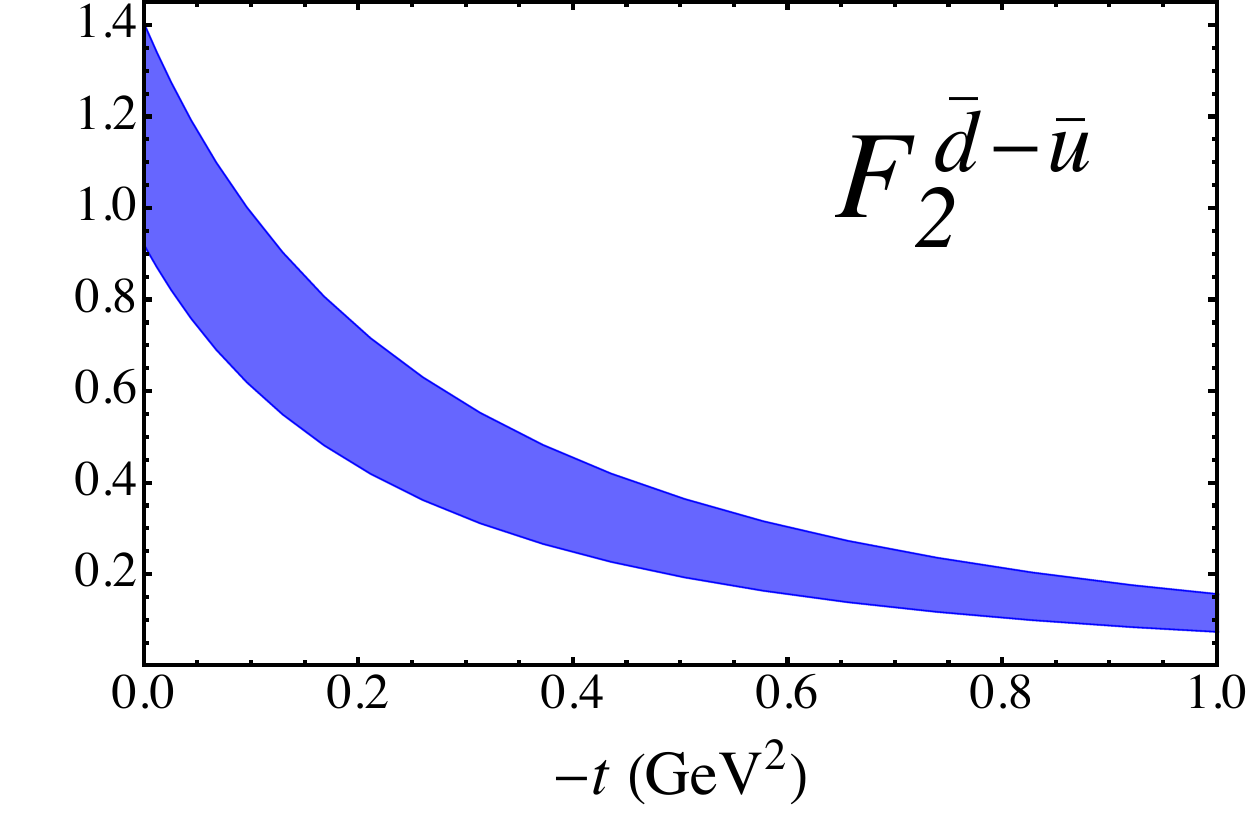} 
   \vspace{0pt}
\end{minipage}  
\\[-0.4cm]
\begin{minipage}[t]{.45\linewidth}
\hspace*{-0.5cm}\includegraphics[width=1.12\textwidth, height=5.5cm]{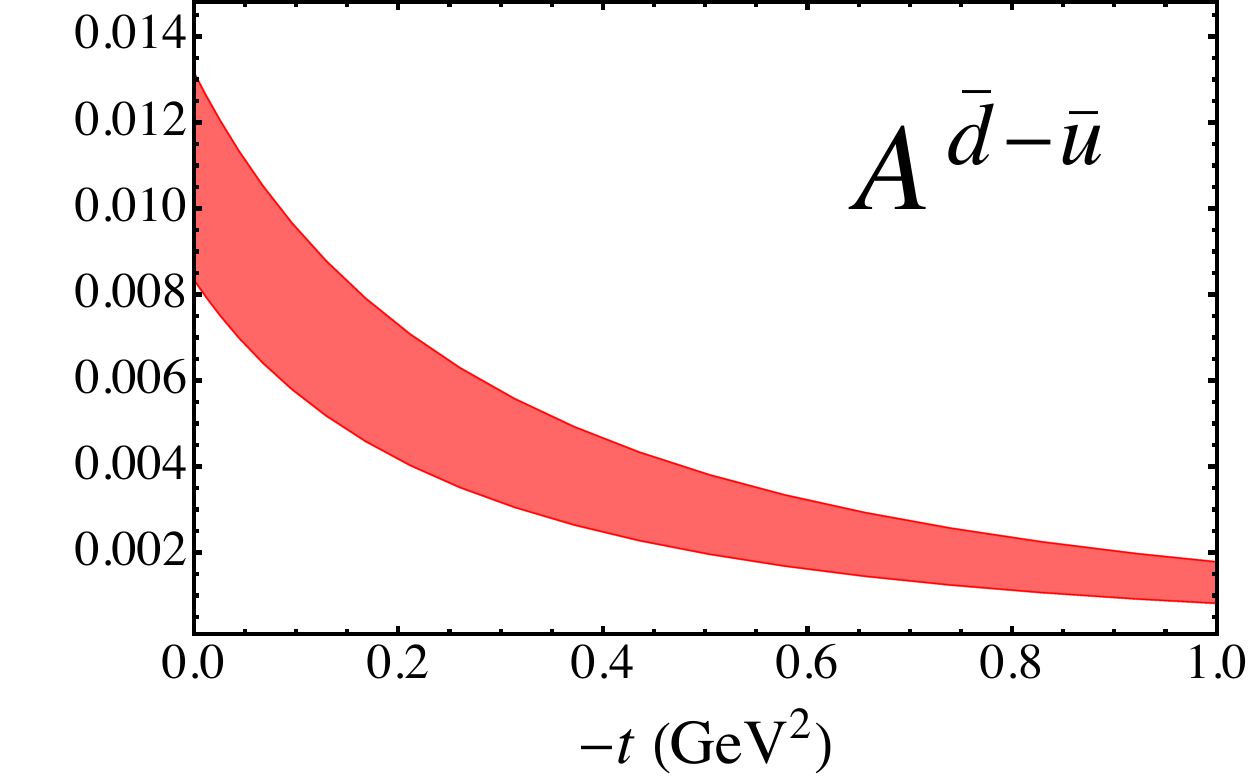}   
 \vspace{0pt}
\end{minipage}
\hfill
\begin{minipage}[t]{.45\linewidth}   
\hspace*{-0.85cm} \includegraphics[width=1.1\textwidth, height=5.5cm]{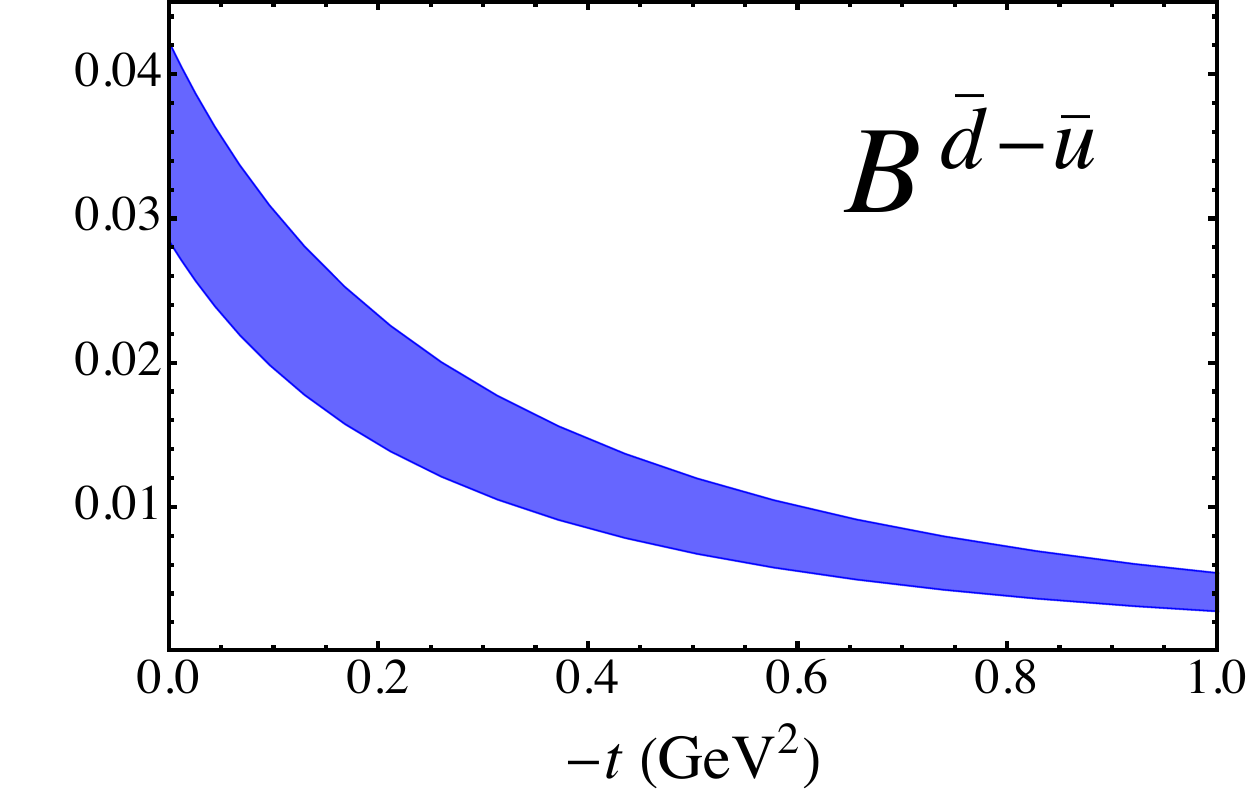}  
   \vspace{-12pt}
\end{minipage} 
\caption{Meson loop contributions to the Dirac $F_1^q$ and Pauli $F_2^q$ (upper panels) and gravitational form factors $A^q$ and $B^q$ (lower panel) versus $-t$ for $q=\bar{d}-\bar{u}$.}
\label{fig:FFs_dubar}
\end{figure}

In Fig.~\ref{fig:FFs_dubar} we illustrate the $\bar{d}-\bar{u}$ contributions to the Dirac $F_1^q$, Pauli $F_2^q$, and gravitational $A^q$ and $B^q$ form factors.
For the pion loop coupling diagrams in Fig.~\ref{fig:diagrams}, these are given in terms of the moments of the $H_u$ and $E_u$ GPDs at zero skewness by
\begin{subequations}
\begin{eqnarray}
F_1^{\bar{d}-\bar{u}}(t)
&=& \int_0^1 \dd{x} \big( H_u(x,0,t)+H_u(-x,0,t) \big), 
\\
F_2^{\bar{d}-\bar{u}}(t)
&=& \int_0^1 \dd{x} \big( E_u(x,0,t)+E_u(-x,0,t) \big), 
\\
A^{\bar{d}-\bar{u}}(t)
&=& \int_0^1 \dd{x}\, x \big( H_u(x,0,t)+H_u(-x,0,t) \big),  
\\
B^{\bar{d}-\bar{u}}(t)
&=& \int_0^1 \dd{x}\, x \big( E_u(x,0,t)+E_u(-x,0,t) \big),
\end{eqnarray}
\end{subequations}
where we have used the crossing symmetry and isospin relations discussed in Sec.~\ref{sec:res_GPDs} to relate the GPDs for the $u$ and $d$ flavors.
From the calculated $H^u$ GPD, we find the $\bar{d}-\bar{u}$ contribution to the Dirac form factor at $t=0$ is
    $F_1^{\bar{d}-\bar{u}}(0) = 0.11(2)$.
Note that $F_1^{\bar{d}-\bar{u}}(0)$ corresponds to the light antiquark PDF asymmetry, 
\begin{eqnarray}
F_1^{\bar{d}-\bar{u}}(0)
&=& \int_0^1 \dd{x} \big( \bar{d}(x)-\bar{u}(x) \big),
\end{eqnarray}
which has been the subject of considerable interest in the literature~\cite{HERMES:1998uvc, NuSea:2001idv, SeaQuest:2021zxb, Salamu:2014pka}.
Our result is consistent with phenomenological extractions from experimental Drell-Yan lepton-pair production data in $pp$ and $pd$ collisions,
    $\int_{0.015}^{0.35} \dd{x} (\bar d - \bar u) = 0.0803(11)$ 
\cite{NuSea:2001idv}, as well as from global QCD analyses of all high-energy scattering data~\cite{Cocuzza:2021cbi, Accardi:2023gyr}.
The Pauli form factor, on the other hand, is an order of magnitude larger than the Dirac form factor,
    $F_2^{\bar{d}-\bar{u}}(0) = 1.15(25)$,
which is consistent with previous calculations of meson loop effects in chiral effective theory~\cite{He:2022leb}.
Although challenging to obtain directly from experimental measurements, this contribution may be accessible in future lattice QCD calculations.

For the $\bar d - \bar u$ contributions to the gravitational form factors, we find in the forward limit 
    $A^{\bar{d}-\bar{u}}(0) = 0.01_{-0.002}^{+0.003}$ and 
    $B^{\bar{d}-\bar{u}}(0) = 0.035(7)$, 
which is consistent with the previous results from Ref.~\cite{He:2022leb}. 
Note that the $A^{\bar{d}-\bar{u}}$ form factor represents the $\bar d - \bar u$ asymmetry for the light antiquark momentum fractions, 
\begin{eqnarray}
A^{\bar{d}-\bar{u}}(0)
&=& \int_0^1 \dd{x} x \big( \bar{d}(x)-\bar{u}(x) \big).
\end{eqnarray}
Our calculated result can be compared with the truncated moment obtained by the SeaQuest Collaboration from $pp$ and $pd$ Drell-Yan data at Fermilab~\cite{FNALE906:2022xdu},
    $\int_{0.13}^{0.45} \dd{x}\, x \big( \bar{d}(x)-\bar{u}(x) \big)
    \approx 0.00318$. 
The numerical value for the form factor $B^{\bar{d}-\bar{u}}(0)$ is about 3 times larger than the value of $A^{\bar{d}-\bar{u}}(0)$, and is also consistent with the prediction in the large-$N_c$ limit~\cite{Goeke:2001tz}.

\section{Summary}
\label{sec.summary}

Within the framework of nonlocal chiral effective theory, we have calculated the one-loop contributions to the spin-averaged GPDs of the proton at nonzero skewness $\xi$ from coupling to pseudoscalar meson loops with intermediate octet and decuplet baryon states.
We derived the generalization of convolution formulas for nonzero skewness, which, unlike the zero skewness case, now involve both the DGLAP and ERBL regions.
We verified that the moments of the generalized convolution formulas for the GPDs satisfy the correct polynominal properties.

To regularize the ultraviolet divergences in the loop integrations for the GPDs, we introduced a relativistic regulator derived self-consistently from the nonlocal Lagrangian.
Using constraints for this regulator obtained from previous analyses of meson loop contributions to PDFs and GPDs~\cite{He:2017viu, He:2018eyz, He:2022leb}, in addition to phenomenological input for the pion GPD~\cite{Amrath:2008vx} and distribution amplitude~\cite{Teryaev:2001qm, Kivel:2002ia, Diehl:2003ny}, we studied the detailed dependence of the $H_q$ and $E_q$ GPDs for the $u$ and $d$ quarks on the parton momentum fraction $x$, skewness $\xi$, and four-momentum transfer squared $t$.
At the kinematics considered in this analysis, we generally find that the absolute values of the Pauli GPD $E_q$ are significantly larger than those of the Dirac GPD $H_q$, with both suppressed for increasing $-t$.

For the $\bar d-\bar u$ flavor combination, we computed the contribution from pion loops to the Dirac and Pauli electromagnetic and the gravitational form factors as a function of $t$.
At $t=0$ the normalization of the $F_1^{\bar d-\bar u}$ form factor is equivalent to the $x$-integrated value of the $\bar d-\bar u$ PDF asymmetry, and for the gravitational $A^{\bar{d}-\bar{u}}(0)$ form factor the normalization corresponds to the $x$-weighted moment of $\bar d-\bar u$.
Our results are consistent with phenomenological extractions of both asymmetries from experimental data~\cite{NuSea:2001idv, Cocuzza:2021cbi}, and with previous zero skewness meson loop GPD calculations~\cite{He:2022leb}.

An obvious extension of the present work is the inclusion of contributions from the coupling to intermediate state baryons, similar to the zero skewness GPD analysis in Ref.~\cite{He:2022leb}.
This will allow comparison with quark as well as antiquark GPDs, as would be needed for studies of strange quark contributions to proton GPDs, for instance~\cite{Salamu:2018cny, Salamu:2019dok}, or gravitational form factors.
Such a strategy would also allow the calculation of chiral loop contributions to spin-dependent GPDs, and even generalized transverse momentum distributions, which will provide further insights into the three-dimensional structure of the proton.

\acknowledgments

We thank Yuxun Guo and Shuzhe Shi for helpful discussions. F.H. is  supported by the National Science Foundation under Award PHY-1847893. This work was supported by NSFC under Grant No.~11975241, the DOE Contract No.~DE-AC05-06OR23177, under which Jefferson Science Associates, LLC operates Jefferson Lab, DOE Contract No.~DE-FG02-03ER41260, and in part within the framework of the Quark-Gluon Tomography (QGT) Topical Collaboration, under Contract No.~DE-SC0023646.

\appendix
\section{Proof of polynomiality of GPDs}
\label{sec.appendix}

Polynomiality requires that the $n$-th moments of GPDs can be expanded as a series in $\xi$,
\begin{eqnarray}
\int_{-1}^1 \dd{x} x^{n-1} \,H_q(x,\xi,t)=\sum_{i=0,\rm even}^{n-1} \left(2\xi\right)^{i} A^{n(i)}_q(t)+(2\xi)^{n}C^{(n)}_q(t)\Big{|}_{n\,\rm even},
\end{eqnarray}
and similarly for the $E_q$.
In analogy with the proof for the form factor in Sec.~\ref{sec.convolution}, the contributions from Eqs.~(\ref{eq:conv_a}), (\ref{eq:conv_b}) and (\ref{eq:conv_d}) can be written as  
\begin{eqnarray}\label{eq:nmomtot12}
\int_{-1}^1 \dd{x} x^{n-1}\,H_q(x,\xi,t)
&=&\int_\xi^1 \dd{y} y^{n-1}\, f(y,\xi,t) \int_{-1}^{1} \dd{\tilde{z}}\, \tilde{z}^{n-1}\, H_{q/\pi}(\tilde{z},\xi/y,t)  \nonumber\\
&&\hspace{-3.5cm}=\int_\xi^1 \dd{y}\,y^{n-1}\, f(y,\xi,t)\left(\sum_{i=0, \rm even}^{n-1} \left(\frac{2\xi}{y}\right)^{i} A^{n(i)}_{q/\pi}(t)+\left(\frac{2\xi}{y}\right)^n C^{(n)}_{q/\pi}(t)\Big{|}_{n\,\rm even}\right).
\end{eqnarray}
The $n$-th moments of GPDs from Eq.~(\ref{eq:conv_c}) can be expressed as
\begin{eqnarray}
\label{eq:region3}
\int_{-\xi}^\xi \dd{x} x^{n-1}\,H_q(x,\xi,t)
&=&\int_{-\xi}^\xi \dd{y}\,\xi^{n-1}\,f(y,\xi,t)\frac{1}{\pi}\frac{\xi}{y} \int_{s_0}^\infty \dd{s}\frac{\text{Im}\int_0^1 \dd{\eta}\,(2\eta-1)^{n-1}\,\Phi_{q/\pi}({\eta,\frac{\frac{y}{\xi}+1}{2},s)}}{s-t+i\epsilon}  
\nonumber\\
&&\hspace{-3cm}=\int_{-\xi}^\xi \dd{y}\,\xi^{n-1}\,f(y,\xi,t)\left(\sum_{i=0,\rm even}^{n-1}  2^i\left(\frac{y}{\xi}\right)^{n-i-1} A^{n(i)}_{q/\pi}(t)+2^{n}\frac{\xi}{y}C^{(n)}_{q/\pi}(t)\Big{|}_{n\,\rm even}\right),
\end{eqnarray} 
where we have used the property of GDAs,
%
\begin{eqnarray}
\label{eq:GDA_pro}
\int_0^1 \dd{\eta}\,(2\eta-1)^{n-1}\,\Phi_{q/\pi}(\eta,z,s)
&=&\frac{2^{n-1}}{(p^++p'^+)^{n}}\langle 0|\bar{q}\gamma^+ (\mathop{i\partial^+}\limits ^{\leftrightarrow})^{n-1} q(0)|\pi(p)\pi(p')\rangle  
\nonumber\\
&&\hspace{-2cm}=\frac{2^{n-1}}{(p'^++p^+)^n}\langle \pi(-p')|\bar{q}\gamma^+ (\mathop{i\partial^+}\limits ^{\leftrightarrow})^n q(0)|\pi(p)\rangle   
\nonumber\\
&&\hspace{-2cm}=\frac{2^{n-1}}{(p'^++p^+)^n}\bigg(\sum_{i=0,\rm even}^{n-1}(p^+-p'^+)(p^++p'^+)^i\left(\frac{p^+-p'^+}{2}\right)^{n-1-i}A_{q/\pi}^{n(i)}(s) \nonumber\\
&&\hspace{2cm} 
 + 2\, (p^++p'^+)^{n}C^{(n)}_{q/\pi}(s)\Big{|}_{n\,\rm even} \bigg)   
\nonumber\\
&&\hspace{-2cm}=\sum_{i=0,\rm even}^{n-1}2^i\left(\frac{p'^+-p^+}{p'^++p^+}\right)^{n-i}A_{q/\pi}^{n(i)}(s)+2^{n}C^{(n)}_{q/\pi}(s)\Big{|}_{n\,\rm even}    \nonumber\\
&&\hspace{-2cm}=\sum_{i=0,\rm even}^{n-1}2^i\left(2z-1\right)^{n-i}A_{q/\pi}^{n(i)}(s)+2^{n}C^{(n)}_{q/\pi}(s)\Big{|}_{n\,\rm even}.
\end{eqnarray} 
The parameter $z$ here is defined as $z=p'^+ / (p^+ + p'^+)$ in the GDA. 
Combining the results in Eq.~(\ref{eq:nmomtot12}) and (\ref{eq:region3}), the $n$-th moments of GPDs can be written
\begin{eqnarray}
\label{eq:nmomtot123}
\int_{-1}^1 \dd{x} x^{n-1}\, H_q(x,\xi,t)
&=&\int_{-\xi}^1 \dd{y}\, y^{n-1}\, f(y,\xi,t)\left(\sum_{i=0,\rm even}^{n-1} \left(\frac{2\xi}{y}\right)^i A^{n(i)}_{q/\pi}(t)
+ \left(\frac{2\xi}{y}\right)^n
C^{(n)}_{q/\pi}(t) \Big{|}_{n\,\rm even}\right)                                      \nonumber\\
&&\hspace*{-2cm}=\sum_{i=0,\rm even}^{n-1}(2\xi)^iA^{n(i)}_{q/\pi}(t)
\int_{-\xi}^1 \dd{y} y^{n-i-1}\,f(y,\xi,t)+(2\xi)^nC^{(n)}_{q/\pi}(t)\int_{-\xi}^1d\,y\,\,\frac{f(y,\xi,t)}{y}    
\nonumber\\
&&\hspace*{-2cm}=\sum_{i=0,\rm even}^{n-1}(2\xi)^iA^{n(i)}_{q/\pi}(t)\left(\sum_{j=0,\rm even}^{n-i-1} (2\xi)^{j} {\cal A}^{n(j)}(t)+\left(2\xi\right)^{n-i}{\cal C}^{(n)}(t)\Big{|}_{n\,\rm even}\right)
\nonumber\\
&& +\, (2\xi)^n\, C^{(n)}_{q/\pi}(t\, ){\cal C'}(t),
\end{eqnarray} 
where ${\cal A}^{n(j)}$, ${\cal C}^{(n)}$, ${\cal C'}$ are the moments of splitting function, they are defined as
\begin{subequations}
\begin{eqnarray}
\int_{-\xi}^1 \dd{y} y^{n-1}\, f(y,\xi,t)
&=& \sum_{j=0,\rm even}^{n-1} (2\xi)^{j} {\cal A}^{n(j)}(t)+\left(2\xi\right)^{n}{\cal C}^{(n)}(t)\Big{|}_{n\,\rm even},  
\\
\int_{-\xi}^1 \dd{y} y^{-1}\, f(y,\xi,t) 
&=& {\cal C'}(t).
\end{eqnarray}
\end{subequations}
Equation~(\ref{eq:nmomtot123}) can be rewritten as
\begin{eqnarray}
\int_{-1}^1 \dd{x} x^{n-1}\, H_q(x,\xi,t)
&=& \sum_{i=0,\rm even}^{n-1}(2\xi)^i \sum_{j=0,\rm even}^{i}A^{n(j)}_{q/\pi}(t){\cal A}^{n(i-j)}(t)
\nonumber\\
&+& (2\xi)^n \left(\sum_{i=0,\rm even}^{n-1} 
A^{n(i)}_{q/\pi}(t)\, {\cal C}^{(n)}(t) + C^{(n)}_{q/\pi}(t)\, {\cal C'}(t)\right).
\end{eqnarray}
Now we obtain the correct polynomiality of GPD $H_q$.
The polynomiality of the magnetic GPD $E_q$ can be obtained by replacing splitting function $f$ by $g$.

\section{Splitting function integrals}
\label{sec.appendixB}

In this appendix, we present the explicit expressions for the integrals of the splitting functions appearing in Sec.~\ref{sec.splitting}, along with the meson-baryon couplings.
For Fig~\ref{fig:diagrams}(a), the splitting functions $f_{\phi B}^{({\rm rbw})}$ and $g_{\phi B}^{({\rm rbw})}$ can be written as
\begin{subequations}
\label{eq:split_rbw_phiB}
\begin{eqnarray}
\label{eq:split_rbw_fphiB}
f_{\phi B}^{({\rm rbw})}(y,\xi,t)
&=&\frac{C_{B\phi}^2}{f^2}
\int\!\frac{\dd[4]{k}}{(2\pi)^4}
\frac{-iF_{\phi B}^{({\rm rbw})}}{D_B(p-k) D_\phi(k+\Delta) D_\phi(k)}
\widetilde{F}(k) \widetilde{F}(k+\Delta)\,
\delta\Big(y+\xi-\frac{k^+}{P^+}\Big),
\notag\\
&& \\
&& \notag\\
g_{\phi B}^{({\rm rbw})}(y,\xi,t)
\label{eq:split_rbw_gphiB}
&=&\frac{C_{B\phi}^2}{f^2}
\int\!\frac{\dd[4]{k}}{(2\pi)^4}
\frac{-iG_{\phi B}^{({\rm rbw})}}{D_B(p-k)D_\phi(k+\Delta) D_\phi(k)}
\widetilde{F}(k) \widetilde{F}(k+\Delta)\,
\delta\Big(y+\xi-\frac{k^+}{P^+}\Big),
\notag\\
&&
\end{eqnarray}
\end{subequations}
where the factors in the numerator of the integrand are given by
\begin{subequations}
\begin{eqnarray}
F_{\phi B}^{({\rm rbw})}&=&
\frac{P^+ y }{2 \left(4 M^2 \xi ^2-\xi ^2 t+t\right)}\bigg\{4 (k\cdot p)^2 \Big[(\xi -1) t-4 M^2 \xi ^2\Big]-4 k\cdot p\, k\cdot p' \Big[4 M^2 \xi ^2+(1+\xi)\Big]
\nonumber\\
&+&2 k \cdot p' \left(2 M \MBbar+k^2\right) \left(4 M^2 \xi ^2+\xi  t+t\right)-2 k\cdot p \Big[2 M M^B \left(4 M^2 \xi ^2-2 \xi ^2 t+\xi  t+t\right)
\nonumber\\
&+&k^2 \left((\xi -1) t-4 M^2 \xi ^2\right)+8 M^4 \xi ^2-4 M^2 \xi ^2 t+2 M^2 \xi  t+2 M^2 t+\xi  t^2+t^2 y\Big]\bigg\}
\\
&& \notag\\
G_{\phi B}^{({\rm rbw})}&=&\frac{2 M P^+ y}{\left(\xi ^2-1\right) t-4 M^2 \xi ^2}\bigg\{4 M (1-\xi) \xi  (k\cdot p)^2-4 M \xi  (\xi +1) k\cdot p\,  k\cdot p'
\nonumber\\
&+&2 M \xi  (\xi +1) k\cdot p' \left(2 M \MBbar+k^2\right)+2 k\cdot p \Big[\left((\xi +1) M^B \left((\xi -1) t-2 M^2 \xi \right)\right)
\nonumber\\
&+&k^2 M (\xi -1) \xi -2 M^3 (\xi +1) \xi -M t \left(-\xi ^2+\xi +y+1\right)\Big]
\nonumber\\
&+&k^2 \Big[M^B \left(4 M^2 \xi ^2-\xi ^2 t+t\right)+4 M^3 \xi ^2+M t \left(-\xi ^2+\xi +y+1\right)\Big]+2 M^2 t \MBbar (\xi +y)
\bigg\}
\nonumber\\
\end{eqnarray}
\end{subequations}
with $\MBbar \equiv M + M_B$.
The coefficients $C^2_{B\phi}$ in Eqs.~(\ref{eq:split_rbw_phiB}) are given from the Lagrangian density in Eq.~(\ref{eq:j1}), 
\begin{subequations}
\begin{eqnarray}
C_{p\pi^0}&=&\frac{g_A^2}{4}, \\
C_{n\pi^+}&=&\frac{g_A^2}{2}.
\end{eqnarray}
\end{subequations}

For the bubble diagram in Fig.~\ref{fig:diagrams}(b), the splitting function $f^{({\rm bub})}_\phi(y,t)$ is given by
\begin{eqnarray} 
\label{eq:f_phi_bub}
f^{({\rm bub})}_\phi(y,\xi,t)
= \frac{iC_{\phi\phi}}{2 f^2}
\int\!\frac{\dd[4]{k}}{(2\pi)^4}
\frac{F_\phi^{({\rm bub})}}{D_\phi(k+\Delta) D_\phi(k)}\,
\widetilde{F}(k+\Delta)\,
\widetilde{F}(k)\,
\delta\Big(y + \xi-\frac{k^+}{P^+}\Big),
\end{eqnarray}
with 
\begin{eqnarray} 
F_\phi^{({\rm bub})}=\frac{y\, P^+\left[4 k \cdot P \left(4 M^2 \xi ^2+t\right)
+ 2 \xi t\, k \cdot \Delta+t^2 (\xi +y)\right]}{4 M^2 \xi ^2-\xi ^2 t+t},
\end{eqnarray}
where the coefficient for the pion loop is given by
\begin{eqnarray}
C_{\pi\pi} &=& \frac12.
\end{eqnarray}
Similarly, for the additional bubble diagram in Fig.~\ref{fig:diagrams}(c), the splitting function $g'^{({\rm bub})}_\phi(y,\xi,t)$ can be expressed as
\begin{eqnarray} 
\label{eq:f_phi_bub}
g'^{({\rm bub})}_\phi(y,\xi,t)
= \frac{iC'_{\phi\phi}}{f^2}
\int\!\frac{\dd[4]{k}}{(2\pi)^4}
\frac{G_\phi^{({\rm bub})}}{D_\phi(k+\Delta) D_\phi(k)}\,
\widetilde{F}(k+\Delta)\,
\widetilde{F}(k)\,
\delta\Big(y + \xi-\frac{k^+}{P^+}\Big).
\end{eqnarray}
with 
\begin{eqnarray} 
G_\phi^{({\rm bub})}=\frac{2 M P^+ y \left[2k \cdot P \left(\xi ^2-1\right) t+ 4k \cdot \Delta M^2 \xi+2 M^2 t (\xi +y)\right]}{\left(\xi ^2-1\right) t-4 M^2 \xi ^2},
\end{eqnarray}
and the coefficients $C'_{\phi\phi}$ is given by
\begin{eqnarray}
C'_{\phi\phi} &=& 2(b_{10}+b_{11}).
\end{eqnarray}

For Fig~\ref{fig:diagrams}(d), the splitting functions $f_{\phi T}^{({\rm rbw})}(y,\xi,t)$ and $g_{\phi T}^{({\rm rbw})}(y,\xi,t)$ can be written as
\begin{subequations}
\label{eq.fg_phiTrbw}
\begin{eqnarray}
\label{eq.f_phiTrbw}
f^{({\rm rbw})}_{\phi T}(y,\xi,t) 
&=& \frac{C_{T\phi}^2}{f^2}
\int\!\frac{\dd[4]{k}}{(2\pi)^4}
\frac{iF_{\phi T}^{({\rm rbw})}}{D_T(p-k)D_\phi(k+\Delta)D_\phi(k)}\,
\widetilde{F}(k+\Delta)\,
\widetilde{F}(k)\,
\delta\Big(y+\xi-\frac{k^+}{P^+}\Big),
\nonumber\\
&& \\
&& \notag\\
\label{eq.g_phiTrbw}
g^{({\rm rbw})}_{\phi T}(y,\xi,t)
&=& \frac{C_{T\phi}^2}{f^2}
\int\!\frac{\dd[4]{k}}{(2\pi)^4}
\frac{iG_{\phi T}^{({\rm rbw})}}{D_T(p-k)D_\phi(k+\Delta)D_\phi(k)}\,
\widetilde{F}(k+\Delta)\,
\widetilde{F}(k)\,
\delta\Big(y+\xi-\frac{k^+}{P^+}\Big),
\nonumber\\
\end{eqnarray}
\end{subequations}
where the factors in the numerator of the integrand are given by
\begin{subequations}
\begin{eqnarray}\label{eq:decf_int}
F_{\phi T}^{({\rm rbw})}
&=&-\frac{P^+ y}{12 M_T^2 \left(4 M^2 \xi ^2-\xi ^2 t+t\right)}
\bigg\{8 (k \cdot p)^2 k \cdot p' \left(4 M^2 \xi ^2 + (1-\xi)t \right)
\nonumber\\
&+&8 (k \cdot p')^2 k \cdot p \left(4 M^2 \xi ^2 + (1-\xi)t \right)
-4 k \cdot p\, k \cdot p' \Big[24 M^3 M_T \xi ^2+16 M^4 \xi ^2+4 M^2 (\xi +1) t
\nonumber\\
&+&2 M M_T \left(2 \xi ^2+6 \xi +3\right) t-t^2 (2 \xi +y+1)\Big]
+ 4 (k \cdot p)^2 \Big[4 M^3 M_T \xi ^2+8 M^4 \xi ^2
\nonumber\\
&+&2 M^2 \left(-2 \xi ^2+3 \xi +1\right) t+M M_T \left(-12 \xi ^2+11 \xi +1\right) t-(\xi -1) t^2\Big]
\nonumber\\
&-&4 M (k \cdot p')^2 (2 M-M_T) \left(4 M^2 \xi ^2+\xi  t+t\right)-2 k \cdot p \Big[2 M^2 t \left(t \left(2 \xi ^2+2 \xi +3 y-1\right)-6 M_T^2 \xi \right)
\nonumber\\
&+&4 k^2 M \left(4 M^3 \xi ^2+M \left(-2 \xi ^2+\xi +1\right) t-3 M_T (\xi -1) \xi  t\right)+32 M^5 M_T \xi ^2
\nonumber\\
&+&4 M^3 M_T \left(-9 \xi ^2+2 \xi +2\right) t+32 M^6 \xi ^2+8 M^4 \left(-3 \xi ^2+\xi +1\right) t
\nonumber\\
&+&M M_T t \left(12 M_T^2 (\xi -1) \xi +t \left(4 \xi ^2+12 \xi +11 y-5\right)\right)-t^3 (\xi +y)
\nonumber\\
&+&3 t^2M_T^2 (\xi -1)\Big]+2 k \cdot p' \Big[32 M^5 M_T \xi ^2+4 M^3 M_T \left(\xi ^2+2 \xi +2\right) t+32 M^6 \xi ^2
\nonumber\\
&-&4 k^2 M \left(4 M^3 \xi ^2-M \left(2 \xi ^2+\xi -1\right) t-3 M_T \xi  (\xi +1) t\right)-2 M^2 t \left(6 M_T^2 \xi +t (2 \xi +y+1)\right)
\nonumber\\
&+&8 M^4 \left(-\xi ^2+\xi +1\right) t+M M_T t \left(t (2 \xi +y+1)-12 M_T^2 \xi  (\xi +1)\right)+3 M_T^2 (\xi +1) t^2\Big]
\nonumber\\
&+&4 k^2 M \Big[16 M^4 M_T \xi ^2+4 M^2 M_T \left(1-4 \xi ^2\right) t+16 M^5 \xi ^2+4 M^3 \left(1-3 \xi ^2\right) t
\nonumber\\
&+&M t^2 \left(2 \xi ^2+\xi +y-2\right)+3 M_T t^2 \left(\xi ^2+\xi +y-1\right)\Big]
\nonumber\\
&+&t^2 \MTbar (\xi +y) \left(8 M^3-2 M \left(6 M_T^2+t\right)+3 M_T t\right)
\bigg\},
\\
&& \notag\\
G_{\phi T}^{({\rm rbw})}&=&\frac{M P^+ y}{3 M_T^2 \left(4 M^2 \xi ^2-\xi ^2 t+t\right)}\bigg\{8 M (\xi -1) \xi  (k \cdot p)^2 k \cdot p'+8 M \xi  (\xi +1) k \cdot p (k \cdot p')^2
\nonumber\\
&-&4 k \cdot p\, k \cdot p' \Big[6 M^2 M_T (\xi +2) \xi +4 M^3 (\xi +1) \xi -M t (2 \xi +y+1)+M_T \left(\xi ^2-1\right) t\Big]
\nonumber\\
&+&4 (k \cdot p)^2 \Big[M^2 M_T (\xi +11) \xi +2 M^3 (\xi +3) \xi -M \left(\xi ^2+\xi -2\right) t-3 M_T \left(\xi ^2-1\right) t\Big]
\nonumber\\
&-&4(k \cdot p')^2  M^2 \xi  (\xi +1) (2 M-M_T)-2 k \cdot p \Big[2 M^3 \left(t \left(-3 \xi ^2+2 \xi +3 y+2\right)-6 M_T^2 \xi \right)
\nonumber\\
&+&k^2 \left(12 M^2 M_T \xi +4 M^3 (\xi +1) \xi -2 M \left(\xi ^2-1\right) t-3 M_T \left(\xi ^2-1\right) t\right)
\nonumber\\
&+&M^2 M_T \left(t \left(-9 \xi ^2+12 \xi +11 y+4\right)-12 M_T^2 \xi \right)+8 M^4 M_T (\xi +1) \xi +8 M^5 (\xi +1) \xi
\nonumber\\
&-&M t \left(t \left(-\xi ^2+\xi +y+1\right)-3 M_T^2 (\xi -1)\right)+M_T \left(\xi ^2-1\right) t \left(3 M_T^2+t\right)\Big]
\nonumber\\
&+&2 k \cdot p' \Big[k^2 \left(12 M^2 M_T \xi -4 M^3 (\xi -1) \xi +2 M \left(\xi ^2-1\right) t+3 M_T \left(\xi ^2-1\right) t\right)
\nonumber\\
&-&2 M^3 \left(6 M_T^2 \xi +t (\xi +2) \xi +y t\right)+M^2 M_T \left(t (\xi +2) \xi +y t-12 M_T^2 \xi \right)
\nonumber\\
&+&8 M^4 M_T (\xi +1) \xi +8 M^5 (\xi +1) \xi +3 M M_T^2 (\xi +1) t-3 M_T^3 \left(\xi ^2-1\right) t\Big]
\nonumber\\
&+&k^2 \Big[16 M^4 M_T \xi ^2+4 M^2 M_T t \left(-4 \xi ^2+3 \xi +3 y+1\right)+16 M^5 \xi ^2
\nonumber\\
&+&4 M^3 t \left(-3 \xi ^2+\xi +y+1\right)+2 M \left(\xi ^2-1\right) t^2+3 M_T \left(\xi ^2-1\right) t^2\Big]
\nonumber\\
&+&M t \MTbar (\xi +y) \left(8 M^3-2 M \left(6 M_T^2+t\right)+3 M_T t\right)\bigg\},
\end{eqnarray}
\end{subequations}
with $\MTbar \equiv M + M_T$ the average of the nucleon and decuplet baryon masses.
The decuplet baryon couplings for the pion case are give by
\begin{subequations}
\begin{eqnarray}
C^2_{\Delta^+\pi^0} &=& \frac{{\cal C}^2}{3}, \\
C^2_{\Delta^0\pi^+} &=& \frac{{\cal C}^2}{6}, \\
C^2_{\Delta^{++}\pi^-} &=& \frac{{\cal C}^2}{2}
\end{eqnarray}
\end{subequations}

\bibliography{ref}

\end{document}